\documentclass[twoside,leqno,twocolumn]{article}

\usepackage{siamproceedings}

\PassOptionsToPackage{table}{xcolor}
\usepackage[subtle]{savetrees}
\usepackage[T1]{fontenc}
\usepackage{amsfonts}
\usepackage{graphicx}
\usepackage{epstopdf}
\usepackage{enumitem}
\ifpdf
  \DeclareGraphicsExtensions{.eps,.pdf,.png,.jpg}
\else
  \DeclareGraphicsExtensions{.eps}
\fi

\newsiamremark{remark}{Remark}
\newsiamremark{hypothesis}{Hypothesis}
\crefname{hypothesis}{Hypothesis}{Hypotheses}
\newsiamthm{claim}{Claim}

\usepackage{amsopn}

\usepackage{booktabs}
\usepackage{xspace}
\usepackage{tikz}
\usepackage{cleveref}
\usepackage{subcaption}
\usepackage{pgfplots}
\pgfplotsset{compat=1.8}
\usepackage{algorithm}
\usepackage{algpseudocode}
\usepackage{wrapfig}
\usepackage{url}
\usepackage{marginnote}
\usepackage{balance}
\usepackage{enumitem}
\setitemize{leftmargin=*,noitemsep,topsep=0pt,parsep=0pt,partopsep=0pt}
\usepackage{svg}
\usepackage[normalem]{ulem}

\newcommand{\para}[1]{\smallskip\noindent\textbf{#1.}}

\newcommand{\defn}[1]{{\textit{\textbf{\boldmath #1}}}\xspace}

\newcommand{\hpdcaddition}[1]{\textcolor{blue}{#1}}

\newcommand{\code}[1]{\texttt{#1}}

\newcommand{\libname}{WarpSpeed\xspace}

\newcommand{\power}{P2HT\xspace}
\newcommand{\Power}{P2HT\xspace}

\newcommand{\meta}{P2HT(M)\xspace}
\newcommand{\Meta}{P2HT(M)\xspace}

\newcommand{\double}{DoubleHT\xspace}
\newcommand{\Double}{DoubleHT\xspace}

\newcommand{\iceberg}{IcebergHT\xspace}

\newcommand{\icebergMeta}{IcebergHT(M)\xspace}

\newcommand{\chaining}{ChainingHT\xspace}
\newcommand{\Chaining}{ChainingHT\xspace}

\newcommand{\cuckoo}{CuckooHT\xspace}

\newcommand{\doubleMeta}{DoubleHT(M)\xspace}
\newcommand{\DoubleMeta}{DoubleHT(M)\xspace}

\newcommand\ColCell[1]{
  \pgfmathparse{#1<50?1:0}  
    \ifnum\pgfmathresult=0\relax\color{white}\fi
  \pgfmathsetmacro\compA{0}      
  \pgfmathsetmacro\compB{#1/100} 
  \pgfmathsetmacro\compC{1}      
  \edef\x{\noexpand\centering\noexpand\cellcolor[\colorModel]{\compA,\compB,\compC}}\x #1
  } 
\newcolumntype{E}{>{\collectcell\ColCell}m{0.4cm}<{\endcollectcell}}  

\begin{document}


\title{WarpSpeed: A High-Performance Library for Concurrent GPU Hash Tables}

\newcommand{\NEU}{Northeastern University}

\author{
  Hunter McCoy\thanks{Northeastern University}%
  \thanks{\email{mccoy.hu@northeastern.edu}}
  \and
  Prashant Pandey\footnotemark[1]%
  \thanks{\email{p.pandey@northeastern.edu}}
}


\date{}
\maketitle
\begin{abstract}

GPU hash tables are increasingly used to accelerate data processing, but their limited functionality restricts adoption in large-scale data processing applications. Current limitations include incomplete concurrency support and missing compound operations such as upserts.

This paper presents \libname, a library of high-performance concurrent GPU hash tables with a unified benchmarking framework for performance analysis. \libname implements eight state-of-the-art Nvidia GPU hash table designs and provides a rich API designed for modern GPU applications. Our evaluation uses diverse benchmarks to assess both correctness and scalability, and we demonstrate real-world impact by integrating these hash tables into three downstream applications.

We propose several optimization techniques to reduce concurrency overhead, including fingerprint-based metadata to minimize cache line probes and specialized Nvidia GPU instructions for lock-free queries. Our findings provide new insights into concurrent GPU hash table design and offer practical guidance for developing efficient, scalable data structures on modern GPUs.

\end{abstract}



\fancyfoot[C]{\footnotesize \raggedleft Copyright \textcopyright\ 2026 by SIAM\\
Unauthorized reproduction of this article is prohibited} 




\pgfplotsset{
  cachePlot/.style={
    small,
    width = \columnwidth,
    height = .5\columnwidth,
    ylabel={Average probes},
    table/col sep=comma,
    xlabel near ticks,
    x label style={at={(0.5,-0.12)},font=\small},
    y label style={at={(-0.1,0.5)},font=\small},
    ytick style={draw=none},
    ytick scale label code/.code = {},
    yticklabel style={ /pgf/number format/fixed },
    ymajorgrids,
    yminorgrids,
    ymin=4,
    minor tick num=1,
    minor grid style={draw=gray!25},
  },
}

\pgfplotsset{
  LFProbePlot/.style={
    small,
    width = 1.1\columnwidth,
    height = .7\columnwidth,
    ylabel={Average probes},
    table/col sep=comma,
    xlabel near ticks,
    x label style={at={(0.5,-0.12)},font=\small},
    y label style={at={(-0.1,0.5)},font=\small},
    ytick style={draw=none},
    ytick scale label code/.code = {},
    yticklabel style={ /pgf/number format/fixed },
    ymajorgrids,
    yminorgrids,
    ymin=0,
    minor tick num=1,
    minor grid style={draw=gray!25},
  },
}

\pgfplotsset{
  LFPlot/.style={
    small,
    width = 1.15\columnwidth,
    height = .9\columnwidth,
    ylabel={Throughput (B/s)},
    table/col sep=comma,
    xlabel near ticks,
    x label style={at={(0.5,-0.12)},font=\small},
    y label style={at={(-0.1,0.5)},font=\small},
    ytick style={draw=none},
    ytick scale label code/.code = {},
    yticklabel style={ /pgf/number format/fixed },
    ymajorgrids,
    yminorgrids,
    ymin=0,
    minor tick num=1,
    minor grid style={draw=gray!25},
    ylabel style={yshift=0.4cm},
  },
}

\pgfplotsset{
  scalingProbePlot/.style={
    small,
    width = \columnwidth,
    height = .5\columnwidth,
    ylabel={Average probes},
    table/col sep=comma,
    xlabel near ticks,
    x label style={at={(0.5,-0.12)},font=\small},
    y label style={at={(-0.1,0.5)},font=\small},
    ytick style={draw=none},
    ytick scale label code/.code = {},
    yticklabel style={ /pgf/number format/fixed },
    ymajorgrids,
    yminorgrids,
    ymin=0,
    minor tick num=1,
    minor grid style={draw=gray!25},
    xlabel style={yshift=-.3cm},
  },
}

\pgfplotsset{
  scalingPlot/.style={
    small,
    width = .9\columnwidth,
    height = .5\columnwidth,
    ylabel={Throughput (M/s)},
    table/col sep=comma,
    xlabel near ticks,
    x label style={at={(0.5,-0.12)},font=\small},
    y label style={at={(-0.1,0.5)},font=\small},
    ytick style={draw=none},
    ytick scale label code/.code = {},
    yticklabel style={ /pgf/number format/fixed },
    ymajorgrids,
    yminorgrids,
    ymin=0,
    minor tick num=1,
    minor grid style={draw=gray!25},
    ylabel style={yshift=0.4cm},
    xlabel style={yshift=-.3cm},
  },
}

\pgfplotsset{
  BGHTProbePlot/.style={
    small,
    width = \columnwidth,
    height = .5\columnwidth,
    ylabel={Average probes},
    table/col sep=comma,
    xlabel near ticks,
    x label style={at={(0.5,-0.12)},font=\small},
    y label style={at={(-0.1,0.5)},font=\small},
    ytick style={draw=none},
    ytick scale label code/.code = {},
    yticklabel style={ /pgf/number format/fixed },
    ymajorgrids,
    yminorgrids,
    ymin=0,
    minor tick num=1,
    minor grid style={draw=gray!25},
    xlabel style={yshift=-.3cm},
  },
}

\pgfplotsset{
  BGHTPlot/.style={
    small,
    width = 1\columnwidth,
    height = .5\columnwidth,
    ylabel={Throughput (M/s)},
    table/col sep=comma,
    xlabel near ticks,
    x label style={at={(0.5,-0.12)},font=\small},
    y label style={at={(-0.1,0.5)},font=\small},
    ytick style={draw=none},
    ytick scale label code/.code = {},
    yticklabel style={ /pgf/number format/fixed },
    ymajorgrids,
    yminorgrids,
    ymin=0,
    minor tick num=1,
    minor grid style={draw=gray!25},
    ylabel style={yshift=0.4cm},
    xlabel style={yshift=-.3cm},
  },
}

\pgfplotsset{
  sawtoothCombinedPlot/.style={
    small,
    width = \columnwidth,
    height =.5\columnwidth,
    ylabel={Throughput (M/s)},
    table/col sep=comma,
    xlabel near ticks,
    x label style={at={(0.5,-0.12)},font=\small},
    ylabel near ticks,
    y label style={at={(-0.1,0.5)},font=\small},
    ytick scale label code/.code = {},
    yticklabel style={ /pgf/number format/fixed },
    ymajorgrids,
    yminorgrids,
    minor tick num=1,
    minor grid style={draw=gray!25},
    every axis plot/.append style={thick},
  },
}

\pgfplotsset{
  graphPlot1/.style={
        small,
        ybar,
        clip=true,
        width = \columnwidth+5pt,
        height =.5\columnwidth,
        bar width=0.12cm,
        ymajorgrids, tick align=inside,
        enlarge x limits={0.15},
        ymin=0.1,
        axis x line*=bottom,
        axis y line*=left,
        y axis line style={opacity=0},
        ylabel={Throughput (M/s)},
       xtick=data,
       legend to name=graphLegend,
       legend pos = outer north east,
  }
}

\pgfplotsset{
  ProbePlot1/.style={
        small,
        ybar,
        clip=true,
        width = \columnwidth,
        height =.5\columnwidth,
        bar width=0.15cm,
        ymajorgrids, tick align=inside,
        enlarge x limits={0.15},
        ymin=0.1,
        axis x line*=bottom,
        axis y line*=left,
        y axis line style={opacity=0},
        ylabel={Average probes},
       xtick=data,
       legend to name=graphLegend,
       legend pos = outer north east,
       point meta=rawy,
       nodes near coords,
       every node near coord/.append style={rotate = 90, anchor = west, font=\tiny,/pgf/number format/fixed relative,/pgf/number format/precision=2
       },
  }
}

\definecolor{rose}{HTML}{B8336A}

\definecolor{turq}{HTML}{48D1CC}

\definecolor{medGreen}{HTML}{027148}

\definecolor{lightGreen}{HTML}{D3F7AD}

\definecolor{khaki}{HTML}{6A5F31}

\definecolor{mintGreen}{HTML}{98FF98}

\definecolor{gold}{HTML}{B59410}

\definecolor{forestGreen}{HTML}{228C22}

\pgfplotsset{
  p2ExtStyle/.style = {color = red, mark = star},
  p2IntStyle/.style = {color=cyan, mark = square*},
  doubleStyle/.style = {color=orange, mark = square*},
  ihtStyle/.style = {color=magenta, mark = oplus*},
  ihtMetaStyle/.style = {color=medGreen, mark = pentagon*},
  chainingStyle/.style = {color=violet, mark = triangle*},
  warpcoreStyle/.style = {color=forestGreen, mark = square*},
  p2MetaStyle/.style = {color=black, mark = diamond*},  cuckooStyle/.style = {color=turq,mark=square*},
  doubleMetaStyle/.style = {color=brown, mark = triangle*},
  bchtStyle/.style = {color=cyan, mark = square*},
  p2bhtStyle/.style = {color=lightGreen, mark = square*},
  slabStyle/.style = {color=gold, mark = square*},
  p2ExtStyleFill/.style = {fill = red},
  p2IntStyleFill/.style = {fill=cyan},
  doubleStyleFill/.style = {fill=orange},
  ihtStyleFill/.style = {fill=magenta},
  ihtMetaStyleFill/.style = {fill=medGreen},
  chainingStyleFill/.style = {fill=violet},
  warpcoreStyleFill/.style = {fill=forestGreen},
  p2MetaStyleFill/.style = {fill=black},
  doubleMetaStyleFill/.style = {fill=brown},
  cuckooStyleFill/.style = {fill=turq},
}

\section{Introduction}

Hash tables are a fundamental component of large-scale data
processing~\cite{black1998graph, PandeyBJP17, facebookGraphs, fastRobustHashingDatabase,
simpleEfficientRobustHashTables, memoryEfficientHashJoins, iceberg, 10.1145/3725424}, and recent
research has focused on developing high-performance hash tables optimized for
GPUs~\cite{dyCuckoo, GelHash, Ashkiani2018, dacHash, BGHT, hofmeyr2020terabase,
GSLIDE}. These GPU hash tables have gained significant attention for
accelerating data processing tasks~\cite{Rui2020, Cao2023, Chrysogelos2019,
Paul2021, Shanbhag2020, Yogatama2022, gao2021scaling}.  
However, despite these advancements, current GPU hash tables still face
limitations in both functionality and performance, restricting their
effectiveness in modern GPU-based data processing applications. As a result,
many GPU applications are forced to rely on designs that are slower, less
space-efficient, and unnecessarily complex to compensate for these
shortcomings.


Existing GPU hash tables operate either in the \emph{static mode} (insert once
and then only query) or the \emph{Bulk Synchronous Parallel (BSP)}
mode\footnote{In BSP, applications are run in phases called \emph{supersteps},
which consist of per-thread local work followed by a final synchronous round of
global communication.}, limiting their usability in workloads such as
\emph{sparse tensor contractions}~\cite{SpartaSparseTensors} that rely on fully
concurrent hash tables, allowing simultaneous inserts and queries to
efficiently represent tensors and accumulate contraction outputs.
Similarly, \emph{databases}~\cite{gpuDatabaseSurvey} require fine-grained
atomic updates to handle concurrent transactions efficiently. In GPU caching
applications, the \textit{aging} performance—how well a hash table performs as
items are inserted and removed when near capacity—is critical, yet existing GPU
hash tables degrade significantly under such conditions. 
Finally, \emph{genomics applications} like \emph{de-novo}
assembly~\cite{hofmeyr2020terabase, 10.1145/3572848.3577507,singletonSieving} and $k$-mer counting~\cite{gpuKmerCounting,
lockfreeKmerCount, hashKmer} require \textit{upserts}, a compound operation
that either inserts a new key or modifies its value if the key already exists.
However, current GPU hash tables~\cite{dyCuckoo, GelHash, Ashkiani2018,
dacHash, BGHT} do not support concurrent upserts, further restricting their
applicability.


As these examples show, three critical shortcomings of existing GPU hash tables
are: (1) the inability to support fully concurrent operations (simultaneous
insertions, queries, and deletions), (2) the inability to support compounding
operations such as upserts, and (3) the inability to consistently offer high
performance for aging workloads.

\para{Our contributions}
We implement \textbf{\libname}, a library of high-performance, concurrent GPU hash tables for Nvidia GPUs. \libname contains eight GPU hash table designs:
\emph{Iceberg~\cite{iceberg}, Power-of-Two-Choice~\cite{twoChoiceHashing},
  Cuckoo\cite{cuckoo}, Double hashing~\cite{doubleHashing} and
Chaining~\cite{chainingPaper}}. Three of these designs also include new GPU-variants with \emph{fingerprint-based metadata} optimization for higher performance. All hash table designs support lock-free queries while locking is employed to synchronize inserts and deletes. 

We also implement a \defn{unified benchmarking framework} along with an API tailored
to the needs of modern GPU applications to extensively
evaluate concurrent GPU hash maps. Based on the benchmarking results, we provide
a detailed analysis focused on identifying hash table designs that are
well-suited for downstream applications involving massive concurrency on GPUs.
Our benchmarking framework also includes empirical evaluation involving three real-world applications: YCSB~\cite{Cooper2010}, caching~\cite{10.5555/3488766.3488810}, and sparse tensor contractions~\cite{SpartaSparseTensors,AthenaSparseTensors}.

We employ an \emph{adversarial benchmark} to verify the correctness of GPU hash tables
under concurrent workloads. Our analysis reveals that some form of external
synchronization mechanism (synchronization not based on the slot in
the hash table) is essential for supporting concurrent insertions and
deletions. Additionally, our benchmarks quantify the performance overhead
associated with incorporating external synchronization in GPU hash tables.

Building on insights from our experimental analysis, we implement optimizations
to minimize the overhead of achieving full concurrency. First, we introduce
\emph{fingerprint-based metadata}, a compact representation of keys that
reduces the number of cache line probes during operations—a technique
previously used in CPU hash tables~\cite{iceberg}. Second, we leverage
\emph{GPU vector loads} to enable lock-free queries, further
enhancing efficiency and scalability.


\para{Why concurrency is hard} Designing concurrent hash tables on the GPU is
\emph{not} trivial. GPUs offer \defn{weak memory consistency}, which
necessitates the use of locks and global acquire-load semantics to support
concurrent operations without race conditions. A memory consistency model
defines the order in which memory operations appear to execute across multiple
threads.   Using an adversarial workload, we empirically show that employing
only atomic operations results in race conditions in existing concurrent GPU
hash tables such as SlabHash~\cite{Ashkiani2018}
(see~\Cref{subsec:adverserial}). These costs are further exacerbated by the
inclusion of 8-byte keys and 8-byte values. These are necessary for many
practical applications but are both slower to process and require wider memory
strides. Furthermore, there is a wide spectrum of hash table designs, and
determining the one that is most amenable to the GPU architecture requires
extensive analysis and evaluation. 

\para{Cost of concurrency}
Achieving concurrency on the GPU incurs a heavy performance cost compared to
\emph{bulk-synchronous parallel (BSP)} hash tables on the order of up to $20\%$
of the overall performance. Every atomic operation incurs a performance hit of
$\sim 50$ million operations per second, making it crucial to minimize the
number of atomic operations needed. Furthermore, the performance of a hash
table is also highly dependent on the layout of the hash table in memory
(bucket size) and the pattern in which threads access data (tile size).
Determining the optimal layout is critical to maximizing the performance. 

%
For example, the cuckoo hash table presented in this paper is a concurrent
implementation of 3-way bucketed cuckoo hashing. With the concurrent features
disabled, the hash table has functionally identical queries to the Bucketed
Cuckoo Hash Table (BCHT) from Awad et al.~\cite{BGHT}. However, by tuning the
bucket and tile size, we can achieve 2.4---3.8$\times$ performance gains over
the bucketed cuckoo hash table (BCHT).

We show that \defn{stability}~\cite{iceberg}, a property that
guarantees that keys will not be moved after insertion, is a necessary
component for producing efficient downstream applications - the ability to
perform compound operations and avoid locking can improve performance by over
10$\times$ without sacrificing correctness. For example, in tensor contraction,
an item stored in the table can be queried and modified without using locks as
inserted items will never be deleted. Without stability, locks must be acquired
for every operation, limiting throughput.

Finally, we show the \defn{probing scheme}, the pattern in which objects are
inserted/queried in the table, and a table's layout and access patterns in GPU
memory can significantly affect performance. For example, by tuning the tile
size of a bucketed \cuckoo, we can achieve over 3$\times$ higher peak
throughput compared to previous cuckoo tables. 



\para{Following are the key takeaways}

\begin{itemize}[leftmargin=*]
    \item \iceberg, \power, and \double are the fastest for insertion
      (21\%+ faster than alternatives), \double is the
      fastest for queries by up to 20\%, and \cuckoo is the fastest for
      deletions by up to 31\%.
    \item Supporting concurrency incurs 1\% --- 25\%
      overhead compared to BSP.
    \item Tile and bucket size choices greatly impact performance— the best configuration is over 1300\%
      faster than the worst configuration.
    \item Vector GPU loads enable lock-free queries with only 1\% overhead
      compared to BSP queries.
    \item Using Metadata scheme improves performance at high load factors
      and more than triples negative query performance.
    \item Metadata-enabled tables perform best under aging and caching
      workloads, with up to 29\% speedup.

    \item \double and \meta are up to 50\% faster on sparse tensor
      contractions due to stability and adaptability to load.
    \item  \double and \doubleMeta, which are optimized for high load factors, achieve the highest performance on YCSB workloads.

\end{itemize}

We uncover several key insights for designing and implementing high-performance
GPU hash tables. These findings guide the development of new optimizations,
such as \emph{lock-free queries} and \emph{fingerprint metadata}, to design
fully concurrent GPU hash tables more efficiently. Our evaluation provides
researchers with a clear understanding of the current GPU hash table landscape
and offers valuable insights that can drive the development of numerous other
GPU-based data structures. \libname is available at \url{https://github.com/saltsystemslab/warpSpeed} and \url{https://zenodo.org/records/17138553}.

\section{Hash table design space}

In this section, we first discuss the parameters that dominate the hash table
performance and then describe the existing hash table designs in light of these
parameters.


\subsection{Performance and parameters}
\label{subsec:performance}
Hash table performance is measured using three criteria: \emph{insertion
speed}, \emph{query speed}, and \emph{space efficiency}. All hash tables face a
three-way trade-off among these performance criteria. Modern hash table designs
strive to minimize space while maximizing speed. These performance criteria
depend on two hash table parameters:

\begin{itemize}[leftmargin=*]
    \item \textbf{Collision-resolution strategy} Collision resolution (or
      \emph{probing scheme}) is the technique to handle situations where two different
      keys map to the same bucket in the hash table. The collision resolution
      scheme impacts both the insertion and query speed. Such schemes include
      linear probing, double hashing, cuckoo hashing, power-of-two-choice
      hashing, front-yard-backyard hashing, and chaining. 
    \item \textbf{Bucket size:} Bucket size determines the number of keys that
      can be stored in a single bucket. Larger buckets help reduce the load
      variance across buckets and improve the maximum achievable load factor
      and space efficiency. To achieve optimal performance, buckets are often
      sized to fit inside one or a small number of consecutive cache lines.
\end{itemize}

\emph{The most important factor in hash-table speed is the number of cache
lines accessed during updates and queries. Additional performance objectives
are concurrency and space efficiency.} Pandey et al.~\cite{iceberg} showed that
hash tables could simultaneously achieve high performance and space efficiency
by optimizing two criteria: \emph{referential stability} and \emph{low
associativity}.

A hash table is said to be \defn{stable} if the position where a key is stored
is guaranteed not to change until either the element is deleted or the table is
resized~\cite{sandersstability, originalstability, KnuthVol3, iceberg}.
Stability offers several desirable properties. For example, stability enables
simpler concurrency-control mechanisms and thus reduces the impact of locking
on  performance.
Stability also allows easier integration in applications.
For example, in sparse tensor contraction, the values matching each key must be
accumulated atomically. Since keys are never moved, threads can query the table
without locks and modify the value in place.

The \defn{associativity} of a hash table is the number of locations where an
element can be stored. For example, 2-way cuckoo hashing has very low
associativity as the key can only be found in two buckets. Associativity
affects the worst-case performance of operations: a negative query in a cuckoo
table requires, at worst, checking two buckets; the same operation in a double
hashing table can take more than 20 probes as every possible location must be
checked.

\subsection{Hash table designs}
\label{sec:table_designs}

Hash tables are broadly classified into two categories: \emph{open addressing}
and \emph{closed addressing (or chaining)}.

\para{Closed addressing} In a closed addressing table, any number of keys can
be mapped to the same bucket. It is the responsibility of the bucket to store
all the keys passed to it. This necessitates that the buckets are capable of
expansion. 

\emph{Chaining-based} hash tables~\cite{chainingPaper} are the most popular
example of closed addressing hash tables. In a chaining hash table, keys are
mapped to a bucket using a hash function. A pointer-based linked list is
employed in each bucket to handle collisions and store multiple keys mapping to
the same bucket. The overhead of maintaining pointers results in
poor space efficiency (33\%)~\cite{iceberg}, and these tables often have high
associativity as the entire linked list for a bucket may be traversed. 
Some chaining hash table implementations size the nodes in the linked lists to
store multiple key-value pairs to improve cache
efficiency~\cite{david2015asynchronized}.


\para{Open addressing} In open addressing, each bucket can hold a fixed number
of keys. During an insertion, if the target bucket is already full, an empty
bucket must be located using a collision-resolution strategy. If no empty
bucket is found then the hash table is full. Open addressing tables have a
secondary parameter called a \emph{probing scheme} that determines the order in
which buckets are probed. 

\textit{Linear probing and double hashing} employ an array of
buckets each containing a single key: linear probing traverses buckets
linearly, while double hashing~\cite{doubleHashing} employs two hash functions
to avoid clustering and maintain performance at high load factors.

\textit{K-way cuckoo hashing}~\cite{cuckoo} hashes keys to any of
\textit{k} buckets via \textit{k} independent hash functions. If all buckets
are full, the new key replaces a key already in a bucket, and the insertion
procedure is restarted for the evicted key. As discussed in
\Cref{subsec:performance}, this makes the table unstable and can limit
concurrent performance.

In \textit{power-of-two-choice hashing}, each key is hashed into
two buckets, and inserted into the least loaded bucket. Buckets are required to
be sized to at least $\omega(\log \log n)$ to achieve high load factor. This
table often implements \textbf{shortcutting} to improve performance at low load
factors: when the fill of the primary bucket is below 75\%, keys are inserted
directly without querying the alternate bucket.

\textit{Iceberg hashing}~\cite{iceberg} employs a front yard-backyard design:
there is a large front yard (90\%) that contains most keys and uses single
hashing. The backyard is small (10\%) and uses power-of-two choice hashing to
handle collisions. This design is stable and highly concurrent, but the front
yard must be larger than $\log n + o(\log n
)$ for the scheme to work,
resulting in frontyard buckets that span multiple cache lines. To mitigate the
cost of traversal, fingerprint-based metadata is used for the front yard.

\section{Programming model for GPU hash tables}

In this section, we describe the layout and semantics of GPU memory and
execution model. We then briefly discuss the existing GPU hash tables and their
limitations.

\begin{figure}
    \centering
    \includesvg[width=\columnwidth]{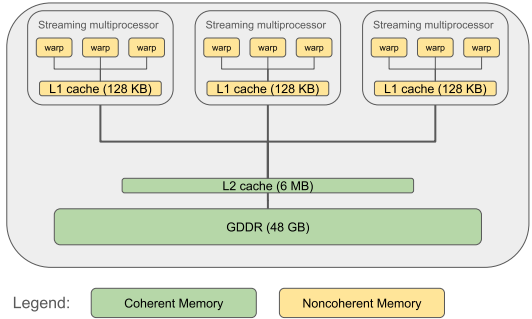}
    \caption{Memory Layout in CUDA. Each streaming multiprocessor executes
    instructions on work groups called warps that share an L1 cache. The warp
  registers and L1 cache are coherent among the threads on a streaming
multiprocessor but are non-coherent with other threads in the GPU. Misses in an
L1 cache trigger a lookup in the coherent L2 cache. Misses in L2 trigger a
lookup in the coherent GDDR, the main memory of the GPU. The non-coherency of
the L1 cache can be bypassed by using \emph{morally strong} loads and stores. 
}
    \label{fig:memory}
\end{figure}

\subsection{GPU memory model} \label{subsec:memory}

A critical component in designing a data structure is the \emph{memory
consistency model}. This model determines how interleaved operations are
visible to each other. For example, in \emph{sequential consistency}, if a read
and write between two different threads are ordered such that the read comes
after the write, the read is guaranteed to observe the write. GPUs follow a
weak memory model known as Relaxed Memory Ordering (RMO)~\cite{ganeshRMO}. In
RMO, reads can be reordered in any fashion, meaning there is no enforced
read-write consistency, as a read can be moved above a paired write.

GPUs follow the RMO model due to their unique internal architecture. A GPU is
composed of units called \defn{streaming multiprocessors (SMs)} that host the
hardware to execute threads. 
Up to 1024 threads can be placed in one SM at a time, with the
threads being organized into work groups of 32 called \emph{warps}. Threads in
a warp share the L1 cache that is \emph{not} coherent with threads outside the
SM.
Threads in an SM share an L1 cache that is \emph{not} coherent with
other L1 caches. All SM L1 caches connect to a coherent L2 cache where updates
are visible to all threads. 
\Cref{fig:memory} shows the layout of GPU memory.

GPUs do not support cache invalidation for the L1 cache: until a cached value
is evicted from L1, it will be reused even if the value has changed in L2 or
GDDR. Even though L2 and GDDR are coherent, reading and writing to these
regions is not guaranteed to be visible to other threads - a cached read in L1
will return a hit and prevent the thread from checking coherent memory.

To fix this issue, PTX ISA 7.8 introduced the notion of \emph{morally strong}
operations~\cite{cudaMemory}.
Morally strong operations define a relationship between operations that
guarantees strong coherency if conditions are met, namely that the instructions
are marked with either "atomic" or "acquire-release" and have a thread-fence
between them.



Since PTX ISA 8.3, GPUs support load and store operations of up to
128-bits through the \code{.b128 type}. This can be combined with acquire and
release tags to allow for coherent loads of up to 16 bytes of data, though it
comes with the restriction that both the read and write must match in size.

\subsection{GPU execution model}
Tasks on a GPU are natively executed in groups called thread blocks of up to
1024 threads. These thread blocks are further subdivided into teams of 32
threads called \defn{warps}. The threads in a warp share underlying hardware,
so performance is maximized when the threads in a warp execute in lockstep or
look at adjacent memory locations simultaneously.

To maximize performance, we group sets of adjacent slots in the hash table into
\defn{buckets}. Each bucket is fully associative: a key that hashes into a
bucket can be found in any of the slots in that bucket. Teams of threads called
\defn{tiles} are used to check several slots in a bucket simultaneously. Tiles
are densely packed into warps: larger tile sizes mean more slots in a bucket
can be queried simultaneously, while smaller tiles lead to better latency
hiding, as more loads are issued per-warp. The optimal tile size thus depends
on the design of the hash table itself and is different for every design. The hash tables do not resize as CUDA does not support reallocation inside of a kernel.


\subsection{A brief history of GPU hash tables}
Alcantara et al.~\cite{realTimeHashingGPU} presented the first GPU
bulk-synchronous parallel (BSP) hash table based on the cuckoo hashing. This
work was improved in the next version~\cite{gpuDatabaseSurvey}, which included
designs for linear probing, quadratic probing, chaining, and a cuckoo hash
table with stashing. Khorasani et al.~\cite{stadiumHashing} then proposed
stadium hashing, a BSP table that allows for simultaneous insertions and
queries. DyCuckoo~\cite{dyCuckoo} is another BSP hash table based on 2-way
bucketed cuckoo hashing.


\para{Warpcore} \label{prev::warpcore} The first table to edge out of the bulk-synchronous framework
was Warpcore from Jünger et al.~\cite{warpcore} which uses a tiling strategy to
efficiently partition warps for operations. This table has a warp-level
implementation and was the first GPU hash table design to utilize cooperative
groups for tiling. Atomics are used to prevent insertion into the same slot,
but the table is explicitly not safe for concurrent operations: the table
offers no guarantee of unique insertion of keys and they explicitly do not
support different operations occurring within the same kernel. Additionally, insertions of key-value pairs are not atomic, making it possible to read a value before it is set.
This table has the ability to perform point operations but is explicitly
\textbf{not safe} for concurrent operations. Correctness is guaranteed when
running the table in a BSP format.

\para{SlabHash} SlabHash~\cite{Ashkiani2018} is a concurrent, dynamic hash
table based on chaining. It uses the atomicCAS to atomically emplace and delete
items from the table, and it maintains a custom allocator called slabAllocator
allocate new buckets.  It does not use locking of any kind when inserting
items. In \cref{subsec:adverserial} we discuss this design in more detail and
highlight the necessity of locking in dynamic hash table designs.

\para{DyCuckoo} DyCuckoo~\cite{dyCuckoo} is a bulk-synchronous
hash table based on 2-way bucketed cuckoo hashing. In this design, called
\textit{2-in-d} hashing, the full hash table is composed of \textit{d}
different subtables. Keys hash to exactly one location inside of a subtable,
and instead of cuckooing between buckets, keys cuckoo between different
subtables. This design allows the table resized incrementally, as a subtable
can be resized without requiring a rehash of the keys in other subtables.
The API for this table is entirely host-controlled.


\para{BGHT} Better GPU Hash Tables (BGHT)~\cite{BGHT} presents
several static hash table designs optimized for use on a GPU. These tables are
all \defn{static}, meaning that keys are inserted during construction of the
table and are never modified. The empirical evaluation in the paper used 32-bit
key-val pairs, but other sizes are supported through the CUDA
\texttt{std::atomic} type. The paper presents static variants of iceberg,
cuckoo, and double hashing tables and finds that the \emph{probe count}, the
number of cache lines touched during an operation, is the main factor that
influences hash table performance.

\para{GELHash} Up to this point, the only fully concurrent hash table for
GPUs in the literature is GELHash. GELHash~\cite{GelHash} is a chaining hash
table that utilizes reader-writer locks adapted from~\cite{GPULocking} to be
fully concurrent. Like SlabHash~\cite{Ashkiani2018}, GELHash uses chaining inside
of buckets, with each bucket being represented by a bucketized linked list.
The table offers two different operation styles, with operations occurring at
either the thread level or the warp level. This is equivalent to using tile
sizes of 1 or 32. A public implementation of this table is not available.

\section{GPU hash table optimizations}

\begin{figure}
    \centering
    \includesvg[width=0.8\columnwidth]{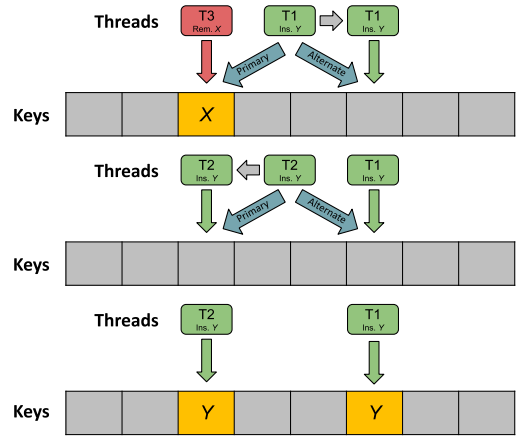}
    \caption{An example of the adversarial scenario with three threads. The
      hash table has an associativity of two. T3 arrives first and tries to
      delete $X$. T1 arrives at the same time and tries to insert $Y$. T1 finds
      the primary bucket occupied and moves to insert $Y$ in the alternate
      bucket. T3 is finished deleting $X$. T2 then arrives and tries to insert
      $Y$. T2 finds the primary bucket empty and tries to insert $Y$ in the
      primary bucket. This sequence of events is a race condition and results
      in duplicate copies of $Y$ in the hash table. Using a lock for each
    bucket or key can avoid the race condition.}
    \label{fig:adversarial}
\end{figure}

In this section, we discuss the challenges and optimizations for implementing a
fully concurrent GPU hash table. We argue that external synchronization is
necessary to correctly support concurrent \emph{upserts} using an adversarial
example. We employ locks in all of our implementations of concurrent GPU hash
tables. Later, we describe how we use metadata (a small fingerprint per key)
technique previously employed by CPU hash tables~\cite{abseil, iceberg} to
speed up GPU hash tables.

\subsection{Supporting concurrent upserts} \label{sec:counter-example}

Existing state-of-the-art concurrent GPU hash tables, such as
SlabHash~\cite{Ashkiani2018}, rely solely on atomic operations for thread
synchronization. They do not employ locks to synchronize among upsert threads,
which results in concurrency bugs.

\emph{Here, we will show that any dynamic and concurrent hash table with
associativity greater than one (which is true for all practical hash tables)
needs some form of external synchronization 
among threads operating on the same key.}

The core issue lies in the lack of external synchronization: when two threads
disagree on the key's location, there is a race condition even if updates are performed atomically.

To see why the above statement is true, consider a hash table with an
associativity of exactly two, meaning each key can be placed in one of two
locations. Any hash table with an associativity greater than two can be
simplified to this scenario by considering only the first two possible
positions for a key.

Now consider the scenario: a key $Y$ has associative locations $L_1$ and $L_2$,
probed in that order during a query/insert, and $L_1$ is currently occupied by
a key $X$. Two threads, $T_1$ and $T_2$ are both attempting to insert key $Y$,
while a third thread, $T_3$, is attempting to remove key $X$ from the hash
table. Without synchronization among the two insertion threads $T_1$ and $T_2$,
there is a potential for conflicting insertion locations, which will, in turn,
result in duplicate keys in the hash table.

This conflict occurs when thread $T_1$ sees $X$ at $L_1$ and moves to check
$L_2$. Meanwhile, thread $T_3$ removes $X$ from $L_1$, and thread $T_2$ finds
$L_1$ available for insertion. Both insertion threads then insert $Y$ into
different locations, resulting in duplicate entries of $Y$. Neither a query
before the insert nor atomicCAS can prevent this conflict. 

To resolve this issue of duplicate keys, threads must establish a serializable
ordering for their operations. In the hash tables discussed in this paper, we
use a lock on the primary bucket of each key to prevent multiple threads from
simultaneously processing the same key. \emph{Concurrent CPU hash table
implementations IcebergHT~\cite{iceberg}, libcuckoo~\cite{libcuckoo}, and
CLHT~\cite{clht} all employ locking to avoid duplicating keys.} 

\para{Alternative approaches}
Locking is not the only method to achieve external synchronization; alternative
approaches include opportunistic concurrency control (OCC) and rollback.  

In OCC, the entire hash table bucket is copied during insertion, and the
updated bucket is atomically swapped in. If the bucket has changed since it was
last accessed, the swap fails, and the operation is retried with the new
bucket. While effective in traditional indexes with infrequent updates, this
approach is expensive on GPUs due to their weak memory model, requiring all
bucket memory to be read and rewritten with acquire-release semantics.
Furthermore, it incurs additional overheads due to dynamic memory management
and a garbage collector for replaced buckets.  

Rollback, on the other hand, involves scanning for duplicate keys after an item
is inserted. This doubles the read cost of each insert operation and requires
all upsert operations to be commutative, as updates could affect multiple
copies of the same key and must be merged.  

Given the high costs and limitations of these alternatives, locking emerges as
the most practical choice for GPUs.

\para{Adversarial benchmark} \label{subsec:adverserial}
%
%
To verify the correctness of existing tables, we construct an adversarial
benchmark to simulate the scenario described in the previous section. The
adversarial benchmark requires two extensions to each hash table design: a
CPU-side function that returns the number of buckets in the table and a
GPU-side function that returns the first bucket a key hashes to.


In this benchmark, keys are generated from a uniform-random distribution and
mapped to their primary buckets until every bucket in the table has exactly two
keys that map to it. The counterexample from~\Cref{fig:adversarial} is then
replayed in every bucket. If the hash table is correct, each bucket should
contain exactly one copy of the key. 


Slabhash~\cite{Ashkiani2018} exhibits the race condition shown
in~\Cref{fig:adversarial} when using the \texttt{insertPairUnique} function,
which claims to prevent duplicate keys from existing in the hash table. In
testing, we generated the issue in roughly 200 out of a million buckets in the
table. None of the concurrent hash table implementations showcased in our
evaluation fail this benchmark as they employ a lock in the primary bucket to
synchronize.

Warpcore~\cite{warpcore} does not have the same race condition as it does not fully implement deletions: the table can not replace tombstone keys. However, this design does suffer from other concurrency issues as discussed in~\Cref{prev::warpcore}.


\begin{figure}
    \centering
    \includesvg[width=0.75\columnwidth]{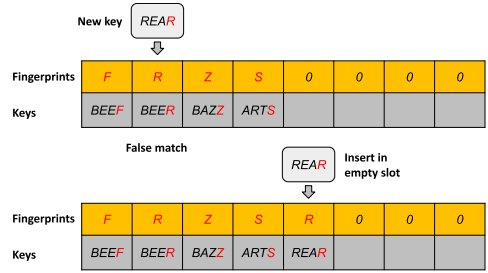}
    \caption{The fingerprint of a key is its last character. To insert a new
    key \textit{REAR}, we first check the metadata for a fingerprint match. If
  that results in a false positive, then insert the new key in the first
available empty slot. The fingerprint is set first in the metadata, and then
the key and value are inserted.}
    \label{fig:md_insertion}
    \vspace{-2em}
\end{figure}


\subsection{Lock-free concurrent queries}
We will now describe how queries can be made entirely lock-free
while being concurrent, significantly improving their performance.

If a key-value pair is written into GPU memory via a morally-strong
128-bit write, then a future morally strong 128-bit read will observe the
key-value pair correctly. \emph{Lock-free concurrent reads are safe only if
the associated write is performed correctly.}


To avoid seeing the key-value pair in a half-written state, the
vector insertions first atomically writes a reservation marker to claim the
slot. This both stops other insertions from writing to the slot and prevents
incorrect data from being read during a query as the key has not been inserted
yet.
After the slot is reserved, a morally strong write is issued by
using a vector store-release instruction to safely write both the key and value
simultaneously.

\subsection{Metadata for reducing cache line probes} \label{sec:metadata}
%
The number of probes required for an operation dominates the
performance in GPU hash tables. This is because acquire-load semantics are
required for all concurrent operations on GPUs as cache invalidation is not
present (weak cache consistency, see~\Cref{subsec:memory}). 
This makes hash table probes significantly more expensive, with overheads
between 20\%-50\% for bucket loads. in the hash table.  when enabling these
instructions. 
This problem is especially exacerbated at high load factors when the number of
probes are significantly high.
Reducing the number of probes required for an operation can improve the
performance of the operation at high load factors and accelerate the table.
The first hash tables to use metadata were Abseil~\cite{abseil} and Iceberg
hashing from Bender et al.~\cite{iceberg}.

In order to reduce the number of probes, we use \emph{metadata}.
\defn{Metadata} refers to a small, lossy fingerprint of a key called a
\emph{tag}. The use of metadata in hash tables was first proposed by
Abseil~\cite{abseil} as a CPU hash table and later employed by
IcebergHT~\cite{iceberg}. Metadata is maintained for each bucket so that
searches within a bucket usually examine at most a single slot within that
bucket. The tags are sized so that the metadata for each bucket is
maintained in a single cache line. \Cref{fig:md_insertion} shows an example of
insertion into a metadata bucket.

The use of 16-bit metadata increases the space cost of each key by 16 bits.
\emph{It does not introduce error or loss in the information as all metadata
matches are compared against the full key.}

Using one 16-bit metadata tag for each key and a bucket size of 32, the total
size of the metadata bucket is 64 bytes, occupying exactly half of a cache line
in a GPU. This is the load size used by the L2 cache, so no bandwidth is wasted
during the load of the metadata. As the tags are lossy, false positive
collisions are possible, but the probability is negligible at $1/65,536$. 
Barring false positives, the use of metadata in power of two choice hashing improves the
positive query cost from an average of 4 to an average of 2.5 bucket probes and
the cost of negative queries from 8 to 2 probes. 

The only downside of using metadata is that insertion and deletion operations
both require an extra write, as the metadata must be updated separately from
the actual key. 
At low bucket load factors, using metadata is more expensive due to the extra writes. However, at
high load factors, these operations are accelerated, as when operating on a
bucket, only one cache line needs to be probed before inserting/deleting a key.

%
%


\section{Implementation}
\label{sec:impl}

All hash tables are written in CUDA 12.2. The CUDA cooperative groups library
is used to partition the warp from a team of 32 into smaller teams called
\emph{tiles} that perform operations inside the table. All hash tables are
templated to accept two different configuration parameters: the \defn{bucket
size} and the \defn{tile size}. The bucket size is the number of key-value
pairs mapped to each bucket, and the tile size is the number of threads in a
warp assigned to the same operation. Tiles are sized to powers of two and are
densely packed into each warp. We evaluate each hash table using every possible
combination of bucket and tile size and employ the one that achieves the
highest performance for the final benchmarks. All tables are configured with
64-bit keys and 64-bit values.
%
%

The following hash table configurations are evaluated:


\begin{itemize}[leftmargin=*]
    \item \textbf{\double:} Each bucket spans across a single cache line (128
      bytes) with 8 key-val (KV) pairs per bucket and 8 threads per tile. The
      metadata variant \doubleMeta uses 4 threads per tile and has buckets of
      size 32.
    
    \item \textbf{\iceberg}: Each bucket spans across 4 cache lines with 32 KV
      pairs per bucket and 8 threads per tile. The front yard (single hashing)
      uses 83\% of the space and the backyard (power-of-two-choice) uses 17\%.
      The metadata variant \icebergMeta uses 4 threads per tile and
      a bucket size of 32.
    
    \item \textbf{\power:} Each bucket spans across 4 cache lines with 32 KV
      pairs per bucket and 8 threads per tile. The metadata variant
      \meta uses 4 threads per tile and a bucket size of 32.
    
    \item \textbf{\chaining:} Each node in the chain spans a single cache line,
      with 7 KV pairs and one pointer per bucket. The table is sized so that
      all chains will have a length of 1 in expectation and is
      initialized with 4 threads per tile. The Gallatin
      allocator~\cite{gallatin} is used to allocate new nodes on the GPU.
    
    \item \textbf{\cuckoo:} Cuckoo hashing is a concurrent implementation of
      the bucketed cuckoo hash table (BCHT) from BGHT~\cite{BGHT}. This table
      has a bucket size of 8 with one cache line per bucket, 4 threads per
      tile, and three-way cuckoo hashing. This implementation uses the
      concurrent insertion strategy from libcuckoo~\cite{libcuckoo}.

\end{itemize}

\subsection{Hash table API}

\begin{itemize}[leftmargin=*]
    \item \textbf{\texttt{bool Upsert(key, value,func *callback):}}
      Upsert~\cite{libcuckoo, iceberg, clht} is the generic implementation of
      both "update" and "insert" and takes three parameters: a key to insert or
      update, a value associated with the key, and a callback function that
      determines what the upsert behavior should be. If the key does not exist,
      it is inserted with the value. If it does exist, the callback function is
      executed to determine the new value. Upsert can implement many different
      operations based on the callback function, including insertions and
      deletions. For example, the function \texttt{f()\{ return; \}} is
      equivalent to "insert if unique" and will not update the value if the key
      already exists, and the function \texttt{f(loc, key,
      val)\{atomicAdd(\&loc->val, val);\}} will atomically accumulate every
      value that matches the key.

    \item \textbf{\texttt{bool Query(key, value):}} Query locates an item in
      the hash table, returning true if it is found. If the key is found, the
      value associated with the key is placed into \texttt{value}.

    \item \textbf{\texttt{bool Erase(key):}} Erase removes a key
      from the hash table, returning true if the key is found. If the key is
    not present, the table is not modified.
    
\end{itemize}

\para{Concurrency and metadata}
For concurrency, all hash tables use one lock bit per bucket, and all the locks
are placed in an external array. We implement metadata variants for
\doubleMeta, \meta, and \icebergMeta. For each key, the lower order 16 bits are
employed as the tag. The metadata for a bucket (32 key-value pairs) fits in
half a cache line and 4 threads per tile are employed for operating on the
metadata.
\defn{Probe count} measures the number of unique cache line accessed by all
threads in a warp during an operation. To calculate the probe count, we measure
the number of unique cache lines accessed by all threads in a warp. The results
shown are the average number of probes per operation.

\addtolength{\tabcolsep}{-0.4em}
\begin{table*}[t]
\centering
    \begin{tabular}{ c | c c c | c c c c | c c c}
    \toprule
    \bf Table & 
    \multicolumn{3}{c}{{\bf Average load probes}} & \multicolumn{4}{| c }{{\bf Average aging probes}} & \multicolumn{3}{| c }{{\bf BSP Query Performance}}\\
    \midrule
    & Insert & Query  & Delete & Insert & Positive & Negative & Delete & Concurrent & Phased & Overhead \\
    & & & & & Query & Query & & (MOps/s) & (MOps/s) & \% \\
    \midrule
    \double & 5.44 & 1.22 & \textbf{4.22} & 77.09 & 1.61 & 80.00 & \textbf{4.57} & \textbf{3166} & 3233 & 2.10\\
    \doubleMeta & \textbf{5.12} & 2.05 & 6.07 & 23.52 & 2.10 & 19.39 & 6.12 & 2624 & 2624 & 0\\
    \iceberg & 	6.53 & 2.97 & 5.97 & 14.85 & 3.12 & 12.02 & 6.09 & 1404 & 1406 & \textbf{0.14} \\
    \icebergMeta & 5.24 & 2.16 & 6.19 & 6.98 & 2.2 & 3.02 & 6.22 & 2598 & 2866 & 9.35\\
    \power & 6.34 & 2.83 & 5.83 & 11.02 & 3.96 & 8.01 & 6.85 & 1471 & 1471 & 0 \\ 
    \meta & 5.25 & 2.13 & 6.15 & \textbf{6.02} & 2.42 & \textbf{2.01} & 6.4 & 2903 & 2909 & .2 \\
    \cuckoo & 9.72 & 3.81 & 4.81 & 10.32 & 4.31 & 8.00 & 5.38 & 2696 & \textbf{3362} & 24.67 \\
    \chaining & 8.13 & \textbf{1.16} & 4.64 & 10.18 & \textbf{1.2} & 2.06 & 4.78 & 2451 & 2468 & .68\\
    BCHT(BGHT) & 1.22* & 2.73 & - & - & - & - & - & - & 1707.79 & -\\
    P2BHT(BGHT) & 2.00* & 1.91 & - & - & - & - & - & - & 656.82 & -\\
    \bottomrule
    \end{tabular}
    \caption{Average number of cache line probes for hash tables. Load probes measure average \#probes per operation as the tables are loaded to 90\% load factor. Aging probes measure average number of probes over 1000 iterations of the aging benchmark. $^*$ values do not use locks, which artificially lowers their probe count.}
    \label{tab:probes}
\end{table*}

\section{Evaluation}
\label{sec:eval}

In this section, we evaluate the performance of various concurrent GPU hash
tables\footnote{\url{https://anonymous.4open.science/r/warpSpeed-9B23/}} (mentioned in~\Cref{sec:impl}) as a function of during load factor. We
also evaluate aging effects, scalability, and space efficiency. We include
state-of-the-art \emph{bulk-synchronous parallel (BSP)} hash tables to measure
the overheads of external synchronization. Our benchmarking framework includes
real-world workloads across multiple domains, including caching systems
inspired by Memcached and Cachelib~\cite{memcached, cachelib}, sparse tensor
contractions relevant to machine learning and scientific computing, and
high-throughput database operations using (YCSB)~\cite{Cooper2010}.

\para{Excluded tables} 
We exclude \textbf{Slabhash}~\cite{Ashkiani2018} and
\textbf{warpcore}~\cite{warpcore} from our concurrent evaluation as they fail the
correctness test for concurrent insertions and deletions.

\para{System specification} All experiments were run on NVIDIA A40 GPU with
48GB of GDDR and 10,752 CUDA cores.

\subsection{Space usage}

Closed-addressing tables suffer from poor space utilization
due to pointer overheads and dynamic allocation compared to open-addressing.

\para{Results} The space usage of all non-metadata open-addressing tables is
identical as these tables all use the same underlying array structure. \double,
\iceberg, \power, and \cuckoo use 16 bytes per key-value pair and can fill to
90\% load factor, leading to space efficiency of 90\%. The metadata tables
\doubleMeta, \icebergMeta, and \meta use an additional two bytes of data per
key-value pair, leading to space efficiency of 80\%. \chaining is not space
efficient due to the overhead of internal pointers and skew in chain length. On
average, \chaining uses 34 bytes per key-value pair for a space efficiency of
42\%.
We omit the table due to space.

\subsection{Concurrency overheads}
Depending on the table design, supporting full concurrency via vector loads incurs an overhead of between 1\% and 25\%
of total performance over BSP implementations.

\para{Setup} The cost of concurrency benchmark measures the overheads of
supporting fully concurrent operations in hash tables compared to the phased
\emph{bulk-synchronous parallel (BSP)} hash tables. We disable the locks and
replace all acquire-release loads with lazy cachable loads to remove the
concurrency overheads. The benchmark is identical to the load factor test, but
all tables are converted to run in a BSP manner. The test is run with 100M
key-value pairs, and we present the aggregate throughput. 
Along with the regular tables, the power-of-two-choice hashing and bucketed
cuckoo hashing from BGHT~\cite{BGHT} are included as a baseline. 

\para{Results} The performance of the BSP hash tables can be seen in the
rightmost section of ~\Cref{tab:probes}. The concurrency overhead is highest in
\cuckoo due to locking for concurrent queries. Most stable
hash table designs have under 1\% difference in performance between concurrent
and phased implementations, with the only exceptions being \double and
\icebergMeta.


The BGHT~\cite{BGHT} tables are slower than their concurrent counterparts as
these tables cannot tune for different tiling strategies. The fastest BGHT
table, the Bucketed Cuckoo Hash table (BCHT), can only achieve peak performance
of 2.2 billion queries per second, despite having a functionally identical
query path to the \cuckoo table.

\begin{figure*}[t]
    \centering
    \ref{lfLegend}
    \begin{subfigure}{.3\textwidth}
\begin{tikzpicture}
  \centering
  \begin{axis}[
    LFPlot,
    enlarge x limits=0.15,
    legend to name=lfLegend,
    legend columns = 6,
    xlabel={Load factor},
    legend style={font=\small},
    y filter/.code={\pgfmathparse{#1*1e-3}\pgfmathresult}
    ]

    \addplot[doubleStyle, x=lf, y=insert, opacity=0.8]   table {data/results/lf/double_hashing.txt};
    \addplot[doubleMetaStyle, x=lf, y=insert, opacity=0.8] table {data/results/lf/double_hashing_metadata.txt};

    \addplot[ihtStyle, x=lf, y=insert,opacity=0.8]   table {data/results/lf/iht_p2_hashing.txt};

    \addplot[ihtMetaStyle, x=lf, y=insert,opacity=0.8]   table {data/results/lf/iht_p2_metadata_full_hashing.txt};

    \addplot[p2ExtStyle, x=lf, y=insert,opacity=0.8]   table {data/results/lf/p2_hashing_external.txt};

    \addplot[p2MetaStyle, x=lf, y=insert,opacity=0.8] table {data/results/lf/p2_hashing_metadata.txt};
 
    \addplot[cuckooStyle, x=lf, y=insert,opacity=0.8]   table {data/results/lf/cuckoo_hashing.txt};

    \addplot[chainingStyle, x=lf, y=insert,opacity=0.8]   table {data/results/lf/chaining_table.txt};



    \addplot[warpcoreStyle, x=lf, y=insert,opacity=0.8]   table {data/results/lf/warpcore.txt};

 
    \legend{\double, \doubleMeta, \iceberg, \icebergMeta, \power, \meta, \cuckoo, \chaining, WarpCore}
  \end{axis}
  \end{tikzpicture}
  \caption{Insertion}
  \label{fig:lf_insert}
\end{subfigure}\hfill
    \begin{subfigure}{.3\textwidth}
\begin{tikzpicture}
  \centering
  \begin{axis}[
    LFPlot,
    table/col sep = comma,
    enlarge x limits=0.15,
    xlabel={Load factor},
    ylabel={},
    y filter/.code={\pgfmathparse{#1*1e-3}\pgfmathresult}
    ]

    \addplot[doubleStyle,opacity=0.8] table[x expr=\thisrowno{0}, y expr=\thisrowno{2}] {data/results/lf/double_hashing.txt};
    \addplot[doubleMetaStyle,opacity=0.8] table[x expr=\thisrowno{0}, y expr=\thisrowno{2}] {data/results/lf/double_hashing_metadata.txt};
    
    \addplot[ihtStyle,opacity=0.8] table[x expr=\thisrowno{0}, y expr=\thisrowno{2}] {data/results/lf/iht_p2_hashing.txt};
    \addplot[ihtMetaStyle,opacity=0.8] table[x expr=\thisrowno{0}, y expr=\thisrowno{2}] {data/results/lf/iht_p2_metadata_full_hashing.txt};

    \addplot[p2ExtStyle,opacity=0.8] table[x expr=\thisrowno{0}, y expr=\thisrowno{2}] {data/results/lf/p2_hashing_external.txt};
    \addplot[p2MetaStyle,opacity=0.8] table[x expr=\thisrowno{0}, y expr=\thisrowno{2}] {data/results/lf/p2_hashing_metadata.txt};
    
    \addplot[cuckooStyle,opacity=0.8] table[x expr=\thisrowno{0}, y expr=\thisrowno{2}] {data/results/lf/cuckoo_hashing.txt};
    
    \addplot[chainingStyle,opacity=0.8] table[x expr=\thisrowno{0}, y expr=\thisrowno{2}] {data/results/lf/chaining_table.txt};

    \addplot[warpcoreStyle,opacity=0.8] table[x expr=\thisrowno{0}, y expr=\thisrowno{2}] {data/results/lf/warpcore.txt};

    

  \end{axis}
  \end{tikzpicture}
  \caption{Query}
  \label{fig:lf_query}
\end{subfigure}
    \begin{subfigure}{.3\textwidth}

\begin{tikzpicture}
  \centering
  \begin{axis}[
    LFPlot,
    table/col sep = comma,
    enlarge x limits=0.15,
    xlabel={Load factor},
    ylabel={},
    y filter/.code={\pgfmathparse{#1*1e-3}\pgfmathresult}
    ]

    \addplot[doubleStyle,opacity=0.8] table[x expr=\thisrowno{0}, y expr=\thisrowno{3}] {data/results/lf/double_hashing.txt};
    \addplot[doubleMetaStyle,opacity=0.8] table[x expr=\thisrowno{0}, y expr=\thisrowno{3}] {data/results/lf/double_hashing_metadata.txt};

    \addplot[ihtStyle,opacity=0.8] table[x expr=\thisrowno{0}, y expr=\thisrowno{3}] {data/results/lf/iht_p2_hashing.txt};
    \addplot[ihtMetaStyle,opacity=0.8] table[x expr=\thisrowno{0}, y expr=\thisrowno{3}] {data/results/lf/iht_p2_metadata_full_hashing.txt};

    \addplot[p2ExtStyle,opacity=0.8] table[x expr=\thisrowno{0}, y expr=\thisrowno{3}] {data/results/lf/p2_hashing_external.txt};
    \addplot[p2MetaStyle,opacity=0.8] table[x expr=\thisrowno{0}, y expr=\thisrowno{3}] {data/results/lf/p2_hashing_metadata.txt};

    \addplot[cuckooStyle,opacity=0.8] table[x expr=\thisrowno{0}, y expr=\thisrowno{3}] {data/results/lf/cuckoo_hashing.txt};
    
    \addplot[chainingStyle,opacity=0.8] table[x expr=\thisrowno{0}, y expr=\thisrowno{3}] {data/results/lf/chaining_table.txt};

    \addplot[warpcoreStyle,opacity=0.8] table[x expr=\thisrowno{0}, y expr=\thisrowno{3}] {data/results/lf/warpcore.txt};


  \end{axis}
  \end{tikzpicture}
  \caption{Deletion}
  \label{fig:lf_remove}
\end{subfigure}

    \caption{Hash table performance for insertions, queries, and deletions as the load factor varies from 0\%--90\%.}
    \label{fig::lf_all}
\end{figure*}
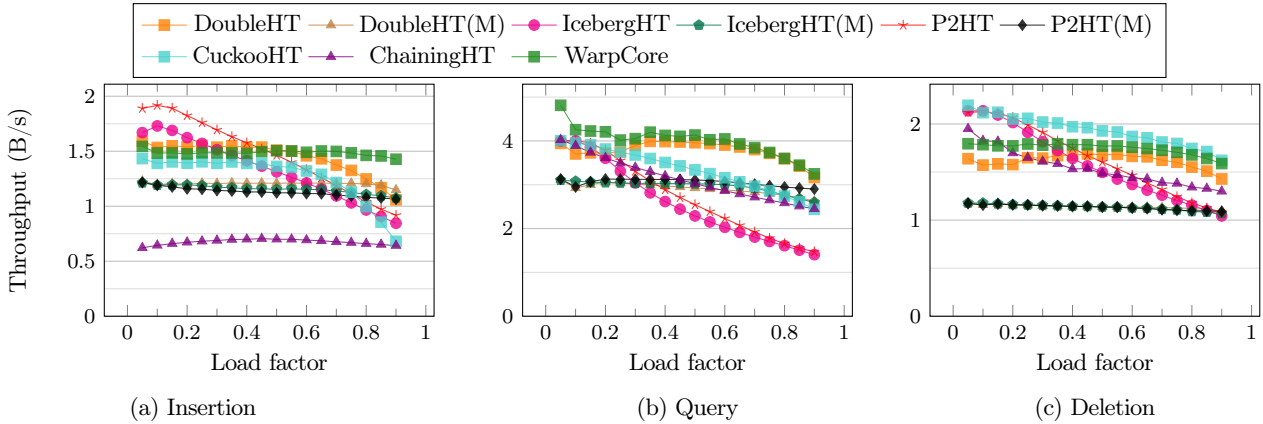

\begin{figure}[ht]


\begin{tikzpicture}
  \centering
  \begin{axis}[
    sawtoothCombinedPlot,
    table/col sep = space,
    enlarge x limits=0.15,
    xlabel={Iteration}
    ]
    \addplot[doubleStyle] table[x=op, y=perf, mark=none]{data/results/sawtooth_combined_1000/double_hashing.txt};
    \addplot[doubleMetaStyle, x=op, y=perf, mark=none] table {data/results/sawtooth_combined_1000/double_hashing_metadata.txt};

    \addplot[ihtStyle] table[x=op, y=perf, mark=none] {data/results/sawtooth_combined_1000/iht_p2_hashing.txt};
    \addplot[ihtMetaStyle] table[x=op, y=perf, mark=none] {data/results/sawtooth_combined_1000/iht_p2_metadata_full_hashing.txt};
    
    \addplot[p2ExtStyle, x=op, y=perf, mark=none] table {data/results/sawtooth_combined_1000/p2_hashing_external.txt};
    \addplot[p2MetaStyle, x=op, y=perf, mark=none] table {data/results/sawtooth_combined_1000/p2_hashing_metadata.txt};
    
    \addplot[cuckooStyle] table[x=op, y=perf, mark=none]{data/results/sawtooth_combined_1000/cuckoo_hashing.txt};
    
    \addplot[chainingStyle] table[x=op, y=perf, mark=none]{data/results/sawtooth_combined_1000/chaining_table.txt};
    

  \end{axis}
  \end{tikzpicture}
  \label{fig:sawtooth_combined}
    \caption{Aggregate results for the aging benchmark when run for 1000 iterations. The graph contains aggregate per-iteration results of the average performance of all operations performed in that iteration.}
    \label{fig:sawtooth_all}
\end{figure}
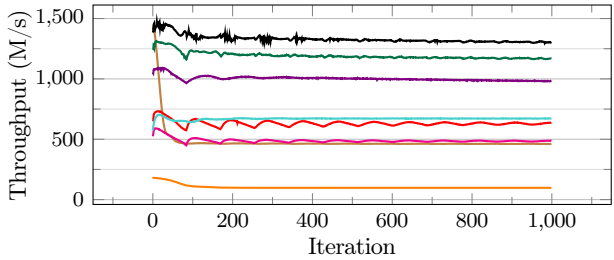



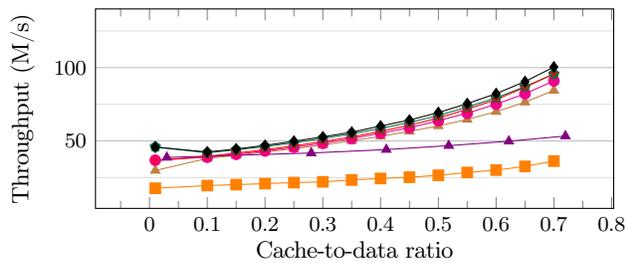
\begin{figure}[ht!]
    \centering

\begin{tikzpicture}
  \centering
  \begin{axis}[
    cachePlot,
    enlarge x limits=0.15,
    legend to name=cacheLegend,
    legend columns = 2,
    xlabel={Cache-to-data ratio},
    table/col sep = space,
    ylabel = {Throughput (M/s)},
    xmax=.70,
    ymax=140,
    ]

    \addplot[doubleStyle] table[x=fill, y=perf]{data/results/cache/double_hashing.txt};
    \addplot[doubleMetaStyle, x=fill, y=perf] table {data/results/cache/double_hashing_metadata.txt};

    \addplot[ihtStyle] table[x=fill, y=perf] {data/results/cache/iht_p2_hashing.txt};
    \addplot[ihtMetaStyle] table[x=fill, y=perf] {data/results/cache/iht_p2_metadata_full_hashing.txt};
    
    \addplot[p2ExtStyle, x=fill, y=perf] table {data/results/cache/p2_hashing_external.txt};
    \addplot[p2MetaStyle, x=fill, y=perf] table {data/results/cache/p2_hashing_metadata.txt};
    
    \addplot[chainingStyle] table[x=fill, y=perf]{data/results/cache/chaining_realigned.txt};
    
    
  \end{axis}
  \end{tikzpicture}
    \caption{Cache performance benchmark. The x-axis is the the ratio of the cache size to the size of the data, with the cache size varying from 1\% of the data size to 100\%. Keys are queried randomly, with the total number of queries being 10$\times$ the number of keys. The y-axis is the throughput in millions of operations per second.}
  \label{fig:cache}
\end{figure}

\subsection{Load benchmark}

Non-metadata tables operate faster during load, but the performance gap closes at high loads. At high load factors,
metadata reduces the number of probes for operations. 

We include WarpCore~\cite{warpcore} only as a baseline as it is not fully concurrent as described in~\Cref{subsec:adverserial}. At 90\% load factor, Warpcore is 24\% faster for insertions, 2\% faster for queries, and 11\% faster for deletions than \double as it does not perform safe synchronization, acquire-release semantics, or atomic key-value pair insertions.

\para{Setup} In the load benchmark, we measure the performance of the hash
tables as they fill from empty to 90\% load factor. The experiment is run with
the hash tables initialized to hold 100M slots. In each iteration, the hash
table is loaded to a set fill percentage ranging from 5\% to 90\%, incrementing
in steps of 5\%, and performance is measured for both insertions and queries at
that fill percentage. For deletions, we remove 5\% of existing keys at a time
until the hash table is empty. We also measure the number of cache line probes
for each operation at different load factors as a representative of the
performance.
%

\para{Insertions} 
Results for insertion are in~\Cref{fig:lf_insert}. For insertions, \power is
the fastest until 35\% load factor with a peak performance of 2.037 billion
insertions per second. The second-fastest table is \iceberg with peak
performance of 2.012 billion inserts per second. 
At small load factors, \power employs a shortcut (see~\Cref{sec:table_designs})
optimization for most insertions, leading to a higher throughput as only one
bucket needs to be loaded.
\iceberg also suffers similar performance degradation as the table must perform
extra loads in the backyard.

Above 35\% load factor, \double has the highest performance, with peak
throughput of 1.76 billion insertions per second. 
Above 65\% load, \cuckoo has the second highest performance of 1.426 billion
insertions per second. However, the performance rapidly decreases at high
load factors, with \cuckoo only performing 804M insertions per second
at 90\% load.
\icebergMeta, \meta, and \doubleMeta have consistent performance for insertions
at all load factors. 

In the load phase, the \power and \iceberg tables feature the lowest probe
counts for insertion, while shortcut optimization is possible. Above 60\% load
factor, the \doubleMeta table has the lowest probe count at 5.12 probes per
insertion, followed closely followed by \icebergMeta at 5.24 probes and \meta at
5.25 probes per insertion.

\para{Queries} Results for queries are in~\Cref{fig:lf_query}. \cuckoo is the
fastest for all loads below 25\% due to its low associativity, with a peak
performance of 4.249 billion queries per second. Above 25\% load factor,
\double is the fastest table with peak performance of 3.96 billion queries per
second.
\power, the second-fastest table below 25\% load, is within 7\% of the
performance of \cuckoo with a peak performance of 3.986 billion queries per
second. 

\Double features the second lowest
average probe count for queries during this benchmark behind \chaining, and \meta and
\icebergMeta always perform a consistent number of loads during queries - as
long as there is not a metadata collision, the \meta loads 1.5 metadata buckets
on average along with one 1 cache line inside of a bucket. With 32 key-value
pairs in a bucket, each bucket spans 4 cache lines. Having to check 3 buckets
in the worst case makes this table very performant when the load is high, and
the \meta table's query performance of 2.605 billion queries per second at 90\%
load factor is the second-fastest of all tables. \Meta is the fastest of
the metadata hash tables with average performance of 2.90 billion queries per
second. 

\para{Deletions} 
Results for deletions are in~\Cref{fig:lf_remove}. \cuckoo is the fastest table
for deletions with a peak performance of 2.17 billion removals per second. This
fast throughput is due to its low associativity and fast worst-case
performance. \power and \iceberg are the second and third-fastest tables for
removal until 40\% load factor, at which point \double is the fastest table
with throughput of 1.77 billion deletions per second. 
\Chaining, \DoubleMeta, \icebergMeta, and \meta have consistent performance for
removals, though \chaining sees a rise in probe count at higher load factors.
The metadata tables maintain consistent performance of 1.17 billion deletions
per second and do not significantly vary for probe count either. 

\begin{figure}[t]
    \centering
    \ref{scalingLegend}\par
    \vspace{0.5cm}
    \begin{subfigure}{\columnwidth}

\begin{tikzpicture}
  \centering
  \begin{axis}[
    scalingPlot,
    xmode=log,
    log basis x={10},
    table/col sep = comma,
    xlabel={\# Items Inserted},
    legend to name=scalingLegend,
    legend columns = 3,
    legend style={font=\small},
    ]

    \addplot[doubleStyle] table[x expr=\thisrowno{0}, y expr=\thisrowno{1}] {data/results/scaling/double_hashing.txt};

    \addplot[doubleMetaStyle] table[x expr=\thisrowno{0}, y expr=\thisrowno{1}] {data/results/scaling/double_hashing_metadata.txt};
    
     \addplot[ihtStyle] table[x expr=\thisrowno{0}, y expr=\thisrowno{1}] {data/results/scaling/iht_p2_hashing.txt};
    \addplot[ihtMetaStyle] table[x expr=\thisrowno{0}, y expr=\thisrowno{1}] {data/results/scaling/iht_p2_metadata_full_hashing.txt};

    \addplot[p2ExtStyle] table[x expr=\thisrowno{0}, y expr=\thisrowno{1}] {data/results/scaling/p2_hashing_external.txt};
    \addplot[p2MetaStyle] table[x expr=\thisrowno{0}, y expr=\thisrowno{1}] {data/results/scaling/p2_hashing_metadata.txt};
    
    \addplot[cuckooStyle] table[x expr=\thisrowno{0}, y expr=\thisrowno{1}] {data/results/scaling/cuckoo_hashing.txt};
    
    \addplot[chainingStyle] table[x expr=\thisrowno{0}, y expr=\thisrowno{1}] {data/results/scaling/chaining_table.txt};
    

    \legend{\double, \doubleMeta, \iceberg, \icebergMeta, \power, \meta, \cuckoo, \chaining}
    
  \end{axis}
  \end{tikzpicture}

  \caption{Insertion}

  \end{subfigure}
    \begin{subfigure}{\columnwidth}
\begin{tikzpicture}
  \centering
  \begin{axis}[
    scalingPlot,
    xmode=log,
    log basis x={10},
    table/col sep = comma,
    xlabel={\# Items Inserted},
    ]

    \addplot[doubleStyle] table[x expr=\thisrowno{0}, y expr=\thisrowno{2}] {data/results/scaling/double_hashing.txt};

    \addplot[doubleMetaStyle] table[x expr=\thisrowno{0}, y expr=\thisrowno{2}] {data/results/scaling/double_hashing_metadata.txt};
    
     \addplot[ihtStyle] table[x expr=\thisrowno{0}, y expr=\thisrowno{2}] {data/results/scaling/iht_p2_hashing.txt};
    \addplot[ihtMetaStyle] table[x expr=\thisrowno{0}, y expr=\thisrowno{2}] {data/results/scaling/iht_p2_metadata_full_hashing.txt};

    \addplot[p2ExtStyle] table[x expr=\thisrowno{0}, y expr=\thisrowno{2}] {data/results/scaling/p2_hashing_external.txt};
    \addplot[p2MetaStyle] table[x expr=\thisrowno{0}, y expr=\thisrowno{2}] {data/results/scaling/p2_hashing_metadata.txt};
    
    \addplot[cuckooStyle] table[x expr=\thisrowno{0}, y expr=\thisrowno{2}] {data/results/scaling/cuckoo_hashing.txt};
    
    \addplot[chainingStyle] table[x expr=\thisrowno{0}, y expr=\thisrowno{2}] {data/results/scaling/chaining_table.txt};
    
  \end{axis}
  \end{tikzpicture}

  \caption{Query}
\end{subfigure}
    \vspace{-0.5cm}
    \caption{Insertion and query performance for hash tales when scaling the size of the table from 10 Million to 1 Billion key-value pairs.}
    \label{fig:scaling_all}
\end{figure}



\subsection{Scaling benchmark}

Insertion performance drops as GPU tables scale due to reduced
L2 hit rate. Query performance does not change.

\para{Setup} This benchmark measures the performance of insertions and queries
as the tables grow from holding 10M to holding 1B key-value pairs. This
benchmark measures raw throughput and probe counts as the tables are filled to
90\% load factor. Additional performance metrics for this benchmark were
captured using the Nsight Compute tool~\cite{nsight}.

\para{Results} 
The results of this benchmark can be seen
in~\Cref{fig:scaling_all}. Most hash tables maintain consistent performance when scaling from 10M to 1B
items, with all hash tables except \cuckoo having less than 20\% variation in
insertion performance as they scale. Query performance is consistent
across sizes. Due to locking, \cuckoo sees a 25\% decrease in query performance
when scaling. 
We omit the figure due to space.

Even though insertion performance suffers for all tables as the size of the
benchmark increases, the per-operation probe counts do not change for any
table. Memory utilization, occupancy, and the activity rate of individual warps
also do not vary as the size of the tables is increased. The only changing
factor shared among the hash tables was a decrease in L2 cache hits due to the
increased size of the table. For example, in \power, memory utilization remains
over 90\% for all sizes, but L2 hit rate drops from 36\% to 27\% when scaling
the table.






\subsection{Aging benchmark}

Negative queries dominate during the aging benchmark and their performance slows down when hash tables do not support an early exit. Therefore, metadata tables age most gracefully as they support cheaper negative queries.

\para{Setup} The aging benchmark measures the performance of hash tables when
they are aged. In the aging benchmark, all hash tables are first filled to 85\%
load factor and then items are inserted and removed in a set pattern to measure
performance as the hash tables age. In each iteration, a new slice of data
equal to 1\% of the total keys is inserted and  oldest 1\% of keys are removed.
Queries are split into positive and negative queries, and a 1\% slice of known
positive and negative keys (not in the hash table) are queried. The benchmark
is run with each hash table initialized to 100M slots (key-value pairs), with
1000 iterations of insertion, query, and removal in each table for aging. All operations per iteration are performed concurrently in the same kernel.

\para{Results} Performance is recorded per iteration, and the results of this
benchmark are available in~\Cref{fig:sawtooth_all}. \meta shows the highest
average performance at 1.35 billion operations per second, followed by
\icebergMeta at 1.25 billion operations. \power, \iceberg, and \cuckoo show
consistent but lower performance in the benchmark. The performance of \double
and \doubleMeta rapidly drops over the first 100 iterations. This performance
drop occurs because the tables become fully saturated, leading to costly
negative queries and insertions as queries cannot exit early. This can be seen in~\Cref{tab:probes} where the cost of upserts and negative queries raises drastically for these tables.

\icebergMeta and \meta are the fastest for insertions and are 28\% faster than
\chaining, the next best table. For positive queries, \doubleMeta is the
fastest at 2.25 billion queries per second. \cuckoo is the fastest table for
deletions, with 1.276 billion deletions per second due to the low
associativity. For negative queries, \meta is the most performant at 2.956
billion queries per second as it has the smallest probe count
(see~\Cref{tab:probes}). The metadata tags avoid probing the hash table
buckets, resulting in a lower probe count. The same is true for \icebergMeta as
only 3 cache lines must be loaded to identify most negative queries.

\subsection{Caching workload}
\label{subsec:caching}

Tables rapidly age during caching: as a result metadata designs out-compete their non-metadata variants.

\para{Setup}
The caching workload models a GPU hash table managing datasets larger than GPU RAM. The hash table resides on the GPU, while a key-value buffer remains on the CPU. Queries first check the GPU hash table; if a key is missing, it is retrieved from the CPU and inserted into the GPU, evicting an entry in FIFO order if necessary. Evicted keys are returned to the CPU.  
The benchmark uses a 100M-item dataset with 1B uniform-random queries generated via OpenSSL's RAND\_BYTES function. It runs multiple times per hash table, varying the table size from 1\% to 70\% of total keys in 5\% increments. A ring queue, set to 85\% of the hash table size, ensures the table’s maximum load factor never exceeds 85\%.

\para{Results} 
\Cref{fig:cache} presents the performance of various hash tables for the caching workload.  
Across all cache-to-data ratios, the \meta table achieves the highest performance due to its high query throughput. Since every item must be queried before use, either negative or positive queries dominate depending on the cache size. As \meta is the fastest for negative queries and the second-fastest for positive queries, it consistently outperforms other tables. \icebergMeta is the second-fastest table, reaching a peak throughput of 95.7M queries per second, while \Power ranks third with 95.6M operations per second. \iceberg and \double perform slightly worse than \power, with \double being slower than \iceberg below a 35\% load factor. Among open addressing tables, \Double is the slowest, peaking at 39M queries per second due to the high prevalence of negative queries in the benchmark.  

The closed addressing scheme of \chaining makes it unsuitable for caching, as it dynamically grows the hash table. For instance, when the cache is initially sized to 10\% of the dataset, the chaining table expands to 28\% over the benchmark's execution. To fairly compare its performance, we track this growth and align its results with the final cache-to-data ratio. Despite this adjustment, the chaining table has the second-lowest performance, reaching 53M operations per second at a 71\% cache ratio. Finally, as \cuckoo is not stable and cannot perform fused operations, it is unable to run this benchmark.

\begin{table}[t]
\centering
\resizebox{\columnwidth}{!}{%
    \begin{tabular}{| c | c c |}
    \toprule
    \bf Table & NIPS 1 Mode (sec)  & NIPS 3 Mode (sec)\\
    \midrule
    \double & 2.020 & 2.503 \\
    \doubleMeta & 2.165 & 2.821 \\
    \chaining & 3.360 & 3.014 \\
    \iceberg & 2.634 & 2.714 \\
    \icebergMeta & 3.150 & 2.595 \\
    \power & 2.483 & 2.469 \\
    \meta & 2.444 & 2.212 \\

    \bottomrule
    \end{tabular}
    }
    \caption{Sparse contraction of the NIPS tensor with itself. The reported times (in seconds) include setup and contraction for dimensions (2) and (0,1,3).}
    \label{tab:sptc}
\end{table}



\subsection{Sparse tensor contraction}

There are no deletions during sparse tensor contraction. This domain knowledge allows for specialized kernels that improve performance by avoiding locks.

\para{Setup} 
We benchmark the performance of hash tables for sparse tensor contraction
(SpTC) using the same data layout and operations as
SPARTA~\cite{SpartaSparseTensors}. This workload is similar to database hash
joins. The input into SPTC is two tensors (X and Y) in COO format, along with a
description of the dimensions to collapse. In the contraction, every nonzero
point in X is matched on the collapsing dimensions to points in Y. To handle
this matching efficiently, a hash table is used to represent the $Y$ matrix.
The output of the contraction is the union of the non-collapsed dimensions of X
and Y and a value that must be accumulated. A second has table uses upsert to
accumulate the values of the output. In SPARTA, each thread has a unique
accumulator table, and all accumulators are merged at the end of the
contraction. As the number of threads on a GPU is too large to support
individual accumulation, we use one concurrent hash table to accumulate the
output tensor.

\para{Results} For comparison against SPARTA, we perform 1-mode and 3-mode contraction on the
NIPS tensor from the FROSTT~\cite{FROSTT} dataset. Results for this benchmark
can be seen in \Cref{tab:sptc}. For the 1-mode contraction, the fastest table
is \double with a total contraction time of 2.02 seconds, over 8$\times$
faster than SPARTA on CPU with 12 threads. For the 3-mode contraction, the
fastest table was \meta with a total contraction time of 2.212 seconds, over
23$\times$ faster than SPARTA on a CPU with 12 threads.

\begin{table}[t]
\centering
\resizebox{\columnwidth}{!}{%
    \begin{tabular}{ c | c | c c c}
    \toprule
    \bf Table & Load & workload A & workload B & workload C\\
    \hline
    \double & 991 & 1654 & 3147 & 3789\\
    \doubleMeta & 914 & 2140 & 2273 & 2543\\
    \iceberg & 774 & 1243 & 1835 & 2075\\
    \icebergMeta & 889 & 1815 & 2168 & 2398\\
    \power & 820 & 1508 & 2011 & 2199\\
    \meta & 852 & 2037 & 2457 & 2705\\
    \cuckoo & 761 & 15 & 19 & 20 \\
    \chaining & 591 & 1713 & 2276 & 2770\\

    \bottomrule
    \end{tabular}
}
    \caption{YCSB performance for A, B, and C workloads. Result reported is aggregate throughput in Mops/s.}
    \label{tab:ycsb}
\end{table}

\subsection{YCSB workload}

Tables that perform well at high load factors perform the best on YCSB workloads. As there is no aging, the performance of tables like \double and \doubleMeta are unaffected.

\para{Setup} For the YCSB~\cite{Cooper2010} benchmarks, each table is run for 512M operations on a universe of 500M keys. The table is initialized with all keys in the universe present before a workload is run. All workloads follow a Zipfian distribution. Workload A is 50\% updates and 50\% queries, workload B is 5\% updates and 95\% queries, and workload C is 100\% queries.

\para{Results} \Cref{tab:ycsb} shows the YCSB results. Hash tables are kept at high load factors and this benefits tables that operate efficiently at high load factors. For workload A, the performance is dominated by update performance. \doubleMeta is the fastest on this workload with throughput of 2,140 Mops/s, followed by \meta at 2,037 Mops/s. For workload B and workload C, query performance dominates. Accordingly, \double is the fastest for both workloads followed by \meta for workload B and \chaining for workload C. Cuckoo does not perform well on any workload due to the lack of stability which requires it to acquire a lock on all operations.









\section{Discussion and conclusion}

We present \libname, a library of high-performance concurrent
GPU hash tables, and use it to implement and evaluate several designs across
diverse workloads. Our study shows that external synchronization is essential
for supporting concurrent insertions and deletions, and we quantify the
associated overheads. To improve performance, we introduce optimizations like
fingerprint-based metadata and GPU-specific instructions for lock-free queries.
These enhancements enable the stable, concurrent, and composable operations
that are critical for modern data processing.



Our evaluation shows that \double and \meta consistently outperform other designs across multiple benchmarks, with metadata variants providing additional benefits in aging workloads and caching scenarios. For workloads below 90\% load factor, \double provides the highest throughput for insertions and queries, and is the second-fastest table for deletions. This performance extends to real-world data, where \double is the fastest for YCSB setup, workload B, and workload C. For aging workloads, \meta is the best and is the fastest table in both the churn and caching benchmark.

While concurrency introduces overhead compared to bulk-synchronous parallel (BSP) paradigms, our optimizations effectively mitigate these costs, ensuring high-throughput operations. These insights contribute to the broader understanding of GPU hash table design and provide a foundation for future research into high-performance, fully concurrent GPU data structures.

\section*{Acknowledgments}
This research is funded in part by
NSF grant OAC 2339521 and 2517201.


\bibliographystyle{siamplain}
\bibliography{ht_bib}

\begin{thebibliography}{10}

\bibitem{realTimeHashingGPU}
{\sc D.~A. Alcantara, A.~Sharf, F.~Abbasinejad, S.~Sengupta, M.~Mitzenmacher, J.~D. Owens, and N.~Amenta}, {\em Real-time parallel hashing on the gpu}, in ACM SIGGRAPH Asia 2009 Papers, SIGGRAPH Asia '09, New York, NY, USA, 2009, Association for Computing Machinery, \url{https://doi.org/10.1145/1661412.1618500}, \url{https://doi.org/10.1145/1661412.1618500}.

\bibitem{ganeshRMO}
{\sc J.~Alglave, M.~Batty, A.~F. Donaldson, G.~Gopalakrishnan, J.~Ketema, D.~Poetzl, T.~Sorensen, and J.~Wickerson}, {\em Gpu concurrency: Weak behaviours and programming assumptions}, in Proceedings of the Twentieth International Conference on Architectural Support for Programming Languages and Operating Systems, ASPLOS '15, New York, NY, USA, 2015, Association for Computing Machinery, p.~577–591, \url{https://doi.org/10.1145/2694344.2694391}, \url{https://doi.org/10.1145/2694344.2694391}.

\bibitem{Ashkiani2018}
{\sc S.~Ashkiani, M.~Farach-Colton, and J.~D. Owens}, {\em A dynamic hash table for the gpu}, in 2018 IEEE International Parallel and Distributed Processing Symposium (IPDPS), Vancouver, British Columbia, Canada, May 2018, IEEE, p.~419–429, \url{https://doi.org/10.1109/ipdps.2018.00052}, \url{http://dx.doi.org/10.1109/IPDPS.2018.00052}.

\bibitem{BGHT}
{\sc M.~A. Awad, S.~Ashkiani, S.~D. Porumbescu, M.~Farach-Colton, and J.~D. Owens}, {\em Better gpu hash tables}, 2021, \url{https://doi.org/10.48550/ARXIV.2108.07232}, \url{https://arxiv.org/abs/2108.07232}.

\bibitem{memoryEfficientHashJoins}
{\sc R.~Barber, G.~Lohman, I.~Pandis, V.~Raman, R.~Sidle, G.~Attaluri, N.~Chainani, S.~Lightstone, and D.~Sharpe}, {\em Memory-efficient hash joins}, Proc. VLDB Endow., 8 (2014), p.~353–364, \url{https://doi.org/10.14778/2735496.2735499}, \url{https://doi.org/10.14778/2735496.2735499}.

\bibitem{10.5555/3488766.3488810}
{\sc B.~Berg, D.~S. Berger, S.~McAllister, I.~Grosof, S.~Gunasekar, J.~Lu, M.~Uhlar, J.~Carrig, N.~Beckmann, M.~Harchol-Balter, and G.~R. Ganger}, {\em The cachelib caching engine: design and experiences at scale}, in Proceedings of the 14th USENIX Conference on Operating Systems Design and Implementation, OSDI'20, USA, 2020, USENIX Association.

\bibitem{cachelib}
{\sc B.~Berg, D.~S. Berger, S.~McAllister, I.~Grosof, S.~Gunasekar, J.~Lu, M.~Uhlar, J.~Carrig, N.~Beckmann, M.~Harchol-Balter, and G.~R. Ganger}, {\em The cachelib caching engine: design and experiences at scale}, in Proceedings of the 14th USENIX Conference on Operating Systems Design and Implementation, OSDI'20, USA, 2020, USENIX Association.

\bibitem{simpleEfficientRobustHashTables}
{\sc A.~Birler, T.~Schmidt, P.~Fent, and T.~Neumann}, {\em Simple, efficient, and robust hash tables for join processing}, in Proceedings of the 20th International Workshop on Data Management on New Hardware, DaMoN '24, New York, NY, USA, 2024, Association for Computing Machinery, \url{https://doi.org/10.1145/3662010.3663442}, \url{https://doi.org/10.1145/3662010.3663442}.

\bibitem{black1998graph}
{\sc J.~R. Black, C.~U. Martel, and H.~Qi}, {\em Graph and hashing algorithms for modern architectures: Design and performance}, in Algorithm Engineering, 2nd International Workshop, {WAE} '92, Saarbr{\"{u}}cken, Germany, August 20-22, 1998, Proceedings, K.~Mehlhorn, ed., Saarbr{\"{u}}cken, Germany, 1998, Max-Planck-Institut f{\"{u}}r Informatik, pp.~37--48.

\bibitem{doubleHashing}
{\sc A.~Bookstein}, {\em Double hashing}, Journal of the American Society for Information Science, 23 (1972), pp.~402--405, \url{https://doi.org/10.1002/ASI.4630230610}, \url{https://doi.org/10.1002/asi.4630230610}.

\bibitem{Cao2023}
{\sc J.~Cao, R.~Sen, M.~Interlandi, J.~Arulraj, and H.~Kim}, {\em Gpu database systems characterization and optimization}, Proceedings of the VLDB Endowment, 17 (2023), p.~441–454, \url{https://doi.org/10.14778/3632093.3632107}, \url{http://dx.doi.org/10.14778/3632093.3632107}.

\bibitem{gpuDatabaseSurvey}
{\sc J.~Cao, R.~Sen, M.~Interlandi, J.~Arulraj, and H.~Kim}, {\em Gpu database systems characterization and optimization}, Proc. VLDB Endow., 17 (2023), p.~441–454, \url{https://doi.org/10.14778/3632093.3632107}, \url{https://doi.org/10.14778/3632093.3632107}.

\bibitem{10.1145/3725424}
{\sc Y.~Chesetti, B.~Shi, J.~M. Phillips, and P.~Pandey}, {\em Zombie hashing: Reanimating tombstones in graveyard}, Proc. ACM Manag. Data, 3 (2025), \url{https://doi.org/10.1145/3725424}, \url{https://doi.org/10.1145/3725424}.

\bibitem{facebookGraphs}
{\sc A.~Ching, S.~Edunov, M.~Kabiljo, D.~Logothetis, and S.~Muthukrishnan}, {\em One trillion edges: graph processing at facebook-scale}, Proc. VLDB Endow., 8 (2015), p.~1804–1815, \url{https://doi.org/10.14778/2824032.2824077}, \url{https://doi.org/10.14778/2824032.2824077}.

\bibitem{Chrysogelos2019}
{\sc P.~Chrysogelos, M.~Karpathiotakis, R.~Appuswamy, and A.~Ailamaki}, {\em Hetexchange: encapsulating heterogeneous cpu-gpu parallelism in jit compiled engines}, Proceedings of the VLDB Endowment, 12 (2019), p.~544–556, \url{https://doi.org/10.14778/3303753.3303760}, \url{http://dx.doi.org/10.14778/3303753.3303760}.

\bibitem{Cooper2010}
{\sc B.~F. Cooper, A.~Silberstein, E.~Tam, R.~Ramakrishnan, and R.~Sears}, {\em Benchmarking cloud serving systems with ycsb}, in Proceedings of the 1st ACM symposium on Cloud computing, SOCC ’10, ACM, June 2010, \url{https://doi.org/10.1145/1807128.1807152}, \url{http://dx.doi.org/10.1145/1807128.1807152}.

\bibitem{david2015asynchronized}
{\sc T.~David, R.~Guerraoui, and V.~Trigonakis}, {\em Asynchronized concurrency: The secret to scaling concurrent search data structures}, in Proceedings of the Twentieth International Conference on Architectural Support for Programming Languages and Operating Systems, ASPLOS '15, New York, NY, USA, 2015, Association for Computing Machinery, p.~631–644, \url{https://doi.org/10.1145/2694344.2694359}, \url{https://doi.org/10.1145/2694344.2694359}.

\bibitem{clht}
{\sc T.~A. David, R.~Guerraoui, T.~Che, and V.~Trigonakis}, {\em Designing ascy-compliant concurrent search data structures}, 2014, \url{https://infoscience.epfl.ch/handle/20.500.14299/109417}.

\bibitem{memcached}
{\sc B.~Fitzpatrick}, {\em Memcached}.
\newblock \url{https://memcached.org/}, 2003.
\newblock Accessed: 2020-11-06.

\bibitem{gao2021scaling}
{\sc H.~Gao and N.~Sakharnykh}, {\em Scaling joins to a thousand gpus.}, in ADMS@ VLDB, 2021, pp.~55--64.

\bibitem{GelHash}
{\sc L.~Gao, Y.~Xu, C.~Xu, R.~Wang, H.~Yang, Z.~Luan, and D.~Qian}, {\em Towards a general and efficient linked-list hash table on gpus}, in 2019 IEEE 21st International Conference on High Performance Computing and Communications; IEEE 17th International Conference on Smart City; IEEE 5th International Conference on Data Science and Systems (HPCC/SmartCity/DSS), Zhangjiajie, Hunan, China, Aug. 2019, IEEE, p.~1452–1460, \url{https://doi.org/10.1109/hpcc/smartcity/dss.2019.00201}, \url{http://dx.doi.org/10.1109/HPCC/SmartCity/DSS.2019.00201}.

\bibitem{abseil}
{\sc Google}, {\em Google's abseil c++ library}.
\newblock \url{https://abseil.io/}, 2017.
\newblock Accessed: 2020-11-06.

\bibitem{originalstability}
{\sc T.~Gunji and E.~Goto}, {\em Studies on hashing part-1: A comparison of hashing algorithms with key deletion}, Journal of Information Processing, 3 (1980), pp.~1--12.

\bibitem{hofmeyr2020terabase}
{\sc S.~Hofmeyr, R.~Egan, E.~Georganas, A.~C. Copeland, R.~Riley, A.~Clum, E.~Eloe-Fadrosh, S.~Roux, E.~Goltsman, A.~Bulu{\c{c}}, et~al.}, {\em Terabase-scale metagenome coassembly with metahipmer}, Scientific reports, 10 (2020), p.~10689.

\bibitem{warpcore}
{\sc D.~Junger, R.~Kobus, A.~Muller, C.~Hundt, K.~Xu, W.~Liu, and B.~Schmidt}, {\em Warpcore: A library for fast hash tables on gpus}, in 2020 IEEE 27th International Conference on High Performance Computing, Data, and Analytics (HiPC), Virtual, Dec. 2020, IEEE, p.~11–20, \url{https://doi.org/10.1109/hipc50609.2020.00015}, \url{http://dx.doi.org/10.1109/HiPC50609.2020.00015}.

\bibitem{fastRobustHashingDatabase}
{\sc K.~Kara and G.~Alonso}, {\em Fast and robust hashing for database operators}, in 2016 26th International Conference on Field Programmable Logic and Applications (FPL), Lausanne, Switzerland, Aug. 2016, IEEE, p.~1–4, \url{https://doi.org/10.1109/fpl.2016.7577353}, \url{http://dx.doi.org/10.1109/FPL.2016.7577353}.

\bibitem{stadiumHashing}
{\sc F.~Khorasani, M.~E. Belviranli, R.~Gupta, and L.~N. Bhuyan}, {\em Stadium hashing: Scalable and flexible hashing on gpus}, in 2015 International Conference on Parallel Architecture and Compilation (PACT), San Francisco, CA, USA, Oct. 2015, IEEE, p.~63–74, \url{https://doi.org/10.1109/pact.2015.13}, \url{http://dx.doi.org/10.1109/PACT.2015.13}.

\bibitem{KnuthVol3}
{\sc D.~E. Knuth}, {\em The art of computer programming, volume 3: (2nd ed.) sorting and searching}, Addison Wesley Longman Publishing Co., Inc., USA, 1998.

\bibitem{libcuckoo}
{\sc X.~Li, D.~G. Andersen, M.~Kaminsky, and M.~J. Freedman}, {\em Algorithmic improvements for fast concurrent cuckoo hashing}, in Proceedings of the Ninth European Conference on Computer Systems, EuroSys '14, New York, NY, USA, 2014, Association for Computing Machinery, \url{https://doi.org/10.1145/2592798.2592820}, \url{https://doi.org/10.1145/2592798.2592820}.

\bibitem{dyCuckoo}
{\sc Y.~Li, Q.~Zhu, Z.~Lyu, Z.~Huang, and J.~Sun}, {\em Dycuckoo: Dynamic hash tables on gpus}, in 2021 IEEE 37th International Conference on Data Engineering (ICDE), Chania, Greece, Apr. 2021, IEEE, p.~744–755, \url{https://doi.org/10.1109/icde51399.2021.00070}, \url{http://dx.doi.org/10.1109/ICDE51399.2021.00070}.

\bibitem{AthenaSparseTensors}
{\sc J.~Liu, D.~Li, R.~Gioiosa, and J.~Li}, {\em Athena: high-performance sparse tensor contraction sequence on heterogeneous memory}, in Proceedings of the ACM International Conference on Supercomputing, ICS ’21, Rome, Italy, June 2021, ACM, p.~190–202, \url{https://doi.org/10.1145/3447818.3460355}, \url{http://dx.doi.org/10.1145/3447818.3460355}.

\bibitem{SpartaSparseTensors}
{\sc J.~Liu, J.~Ren, R.~Gioiosa, D.~Li, and J.~Li}, {\em Sparta: high-performance, element-wise sparse tensor contraction on heterogeneous memory}, in Proceedings of the 26th ACM SIGPLAN Symposium on Principles and Practice of Parallel Programming, PPoPP ’21, Virtual, Feb. 2021, ACM, p.~318–333, \url{https://doi.org/10.1145/3437801.3441581}, \url{http://dx.doi.org/10.1145/3437801.3441581}.

\bibitem{lockfreeKmerCount}
{\sc G.~Marçais and C.~Kingsford}, {\em {A fast, lock-free approach for efficient parallel counting of occurrences of k-mers}}, Bioinformatics, 27 (2011), pp.~764--770, \url{https://doi.org/10.1093/bioinformatics/btr011}, \url{https://doi.org/10.1093/bioinformatics/btr011}, \url{https://arxiv.org/abs/https://academic.oup.com/bioinformatics/article-pdf/27/6/764/48866141/bioinformatics\_27\_6\_764.pdf}.

\bibitem{singletonSieving}
{\sc H.~McCoy, S.~Hofmey, K.~Yelick, and P.~Pandey}, {\em Singleton sieving: Overcoming the memory/speed trade-off in exascale k-mer analysis}, in SIAM Conference on Applied and Computational Discrete Algorithms (ACDA23), 2023, pp.~213--224, \url{https://doi.org/10.1137/1.9781611977714.19}, \url{https://epubs.siam.org/doi/abs/10.1137/1.9781611977714.19}.

\bibitem{10.1145/3572848.3577507}
{\sc H.~McCoy, S.~Hofmeyr, K.~Yelick, and P.~Pandey}, {\em High-performance filters for gpus}, in Proceedings of the 28th ACM SIGPLAN Annual Symposium on Principles and Practice of Parallel Programming, PPoPP '23, New York, NY, USA, 2023, Association for Computing Machinery, p.~160–173, \url{https://doi.org/10.1145/3572848.3577507}, \url{https://doi.org/10.1145/3572848.3577507}.

\bibitem{gallatin}
{\sc H.~Mccoy and P.~Pandey}, {\em Gallatin: A general-purpose gpu memory manager}, in Proceedings of the 29th ACM SIGPLAN Annual Symposium on Principles and Practice of Parallel Programming, PPoPP '24, New York, NY, USA, 2024, Association for Computing Machinery, p.~364–376, \url{https://doi.org/10.1145/3627535.3638499}, \url{https://doi.org/10.1145/3627535.3638499}.

\bibitem{twoChoiceHashing}
{\sc M.~Mitzenmacher, A.~W. Richa, and R.~Sitaraman}, {\em The Power of Two Random Choices: A Survey of Techniques and Results}, Springer US, 2001, p.~255–312, \url{https://doi.org/10.1007/978-1-4615-0013-1_9}, \url{http://dx.doi.org/10.1007/978-1-4615-0013-1_9}.

\bibitem{chainingPaper}
{\sc P.~Nimbe, S.~Frimpong, and M.~Opoku}, {\em An efficient strategy for collision resolution in hash tables}, International Journal of Computer Applications, 99 (2014), pp.~35--41, \url{https://doi.org/10.5120/17411-7990}.

\bibitem{gpuKmerCounting}
{\sc I.~Nisa, P.~Pandey, M.~Ellis, L.~Oliker, A.~Buluc, and K.~Yelick}, {\em Distributed-memory k-mer counting on gpus}, in 2021 IEEE International Parallel and Distributed Processing Symposium (IPDPS), Virtual, May 2021, IEEE, \url{https://doi.org/10.1109/ipdps49936.2021.00061}, \url{http://dx.doi.org/10.1109/IPDPS49936.2021.00061}.

\bibitem{cudaMemory}
{\sc NVIDIA}, {\em Memory consistency model}, 2024, \url{https://docs.nvidia.com/cuda/parallel-thread-execution/index.html\#memory-consistency-model}.
\newblock [Accessed 29-05-2024].

\bibitem{nsight}
{\sc NVIDIA}, {\em Nsight compute}, Jul 2024, \url{https://docs.nvidia.com/nsight-compute/NsightCompute/index.html}.

\bibitem{cuckoo}
{\sc R.~Pagh and F.~Rodler}, {\em Cuckoo hashing}, Journal of Algorithms, 51 (2004), pp.~122--144, \url{https://doi.org/10.1016/j.jalgor.2003.12.002}.

\bibitem{hashKmer}
{\sc T.~C. Pan, S.~Misra, and S.~Aluru}, {\em Optimizing high performance distributed memory parallel hash tables for dna k-mer counting}, in SC18: International Conference for High Performance Computing, Networking, Storage and Analysis, Dallas, Texas, United States, Nov. 2018, IEEE, p.~135–147, \url{https://doi.org/10.1109/sc.2018.00014}, \url{http://dx.doi.org/10.1109/SC.2018.00014}.

\bibitem{GSLIDE}
{\sc Z.~Pan, F.~Zhang, H.~Li, C.~Zhang, X.~Du, and D.~Deng}, {\em G-slide: A gpu-based sub-linear deep learning engine via lsh sparsification}, IEEE Transactions on Parallel and Distributed Systems, 33 (2022), pp.~3015--3027, \url{https://doi.org/10.1109/TPDS.2021.3132493}.

\bibitem{iceberg}
{\sc P.~Pandey, M.~A. Bender, A.~Conway, M.~Farach{-}Colton, W.~Kuszmaul, G.~Tagliavini, and R.~Johnson}, {\em Iceberght: High performance hash tables through stability and low associativity}, Proceedings of the ACM on Management of Data, 1 (2023), pp.~47:1--47:26, \url{https://doi.org/10.1145/3588727}, \url{https://doi.org/10.1145/3588727}.

\bibitem{PandeyBJP17}
{\sc P.~Pandey, M.~A. Bender, R.~Johnson, and R.~Patro}, {\em A general-purpose counting filter: Making every bit count}, in Proceedings of the 2017 {ACM} International Conference on Management of Data, 2017, pp.~775--787.

\bibitem{Paul2021}
{\sc J.~Paul, S.~Lu, B.~He, and C.~T. Lau}, {\em Mg-join: A scalable join for massively parallel multi-gpu architectures}, in Proceedings of the 2021 International Conference on Management of Data, SIGMOD/PODS ’21, Xi'an, Shaanxi, China, June 2021, ACM, p.~1413–1425, \url{https://doi.org/10.1145/3448016.3457254}, \url{http://dx.doi.org/10.1145/3448016.3457254}.

\bibitem{Rui2020}
{\sc R.~Rui, H.~Li, and Y.-C. Tu}, {\em Efficient join algorithms for large database tables in a multi-gpu environment}, Proceedings of the VLDB Endowment, 14 (2020), p.~708–720, \url{https://doi.org/10.14778/3436905.3436927}, \url{http://dx.doi.org/10.14778/3436905.3436927}.

\bibitem{sandersstability}
{\sc P.~Sanders}, {\em Hashing with linear probing and referential integrity}, 2018, \url{https://doi.org/10.48550/ARXIV.1808.04602}, \url{https://arxiv.org/abs/1808.04602}.

\bibitem{Shanbhag2020}
{\sc A.~Shanbhag, S.~Madden, and X.~Yu}, {\em A study of the fundamental performance characteristics of gpus and cpus for database analytics}, in Proceedings of the 2020 ACM SIGMOD International Conference on Management of Data, SIGMOD/PODS ’20, Portland, Oregon, United States, May 2020, ACM, p.~1617–1632, \url{https://doi.org/10.1145/3318464.3380595}, \url{http://dx.doi.org/10.1145/3318464.3380595}.

\bibitem{FROSTT}
{\sc S.~Smith, J.~W. Choi, J.~Li, R.~Vuduc, J.~Park, X.~Liu, and G.~Karypis}, {\em {FROSTT}: The formidable repository of open sparse tensors and tools}, 2017, \url{http://frostt.io/}.

\bibitem{GPULocking}
{\sc Y.~Xu, L.~Gao, R.~Wang, Z.~Luan, W.~Wu, and D.~Qian}, {\em Lock-based synchronization for gpu architectures}, in Proceedings of the ACM International Conference on Computing Frontiers, CF '16, New York, NY, USA, 2016, Association for Computing Machinery, p.~205–213, \url{https://doi.org/10.1145/2903150.2903155}, \url{https://doi.org/10.1145/2903150.2903155}.

\bibitem{Yogatama2022}
{\sc B.~W. Yogatama, W.~Gong, and X.~Yu}, {\em Orchestrating data placement and query execution in heterogeneous cpu-gpu dbms}, Proceedings of the VLDB Endowment, 15 (2022), p.~2491–2503, \url{https://doi.org/10.14778/3551793.3551809}, \url{http://dx.doi.org/10.14778/3551793.3551809}.

\bibitem{dacHash}
{\sc H.~Zhou, D.~Troendle, and B.~Jang}, {\em Dachash: A dynamic, cache-aware and concurrent hash table on gpus}, in 2021 IEEE 33rd International Symposium on Computer Architecture and High Performance Computing (SBAC-PAD), Belo Horizonte – Brazil, Oct. 2021, IEEE, p.~1–10, \url{https://doi.org/10.1109/sbac-pad53543.2021.00012}, \url{http://dx.doi.org/10.1109/SBAC-PAD53543.2021.00012}.

\end{thebibliography}

\end{document}